# Numerical modelling of shock waves and detonation in complex geometries

## Y. V. Nevmerzhitskiy[1] and S. E. Yakush[2]

[1]Moscow Institute of Physics and Technology
Institutskii per. 9, Dolgoprudny, 141700, Moscow Region, Russia
nevmerzhitski_y@mail.ru

[2]A. Yu. Ishlinskii Institute for Problems in Mechanics of the Russian Academy of Sciences
Ave. Vernadskogo 101/1, Moscow, 119526, Russia
yakush@ipmnet.ru

## Содержание

# ABSTRACT

Cumulation of shock snd detonation waves was considered. Computations were carried out by the second-order central-difference scheme. Cumulation of waves in a cone region with characteristic linear size of 1 meter was studied. Pictures of flow in shock and detonation waves during different time moments were obtained as well as time dependences and maximum pressures for different corner angles.

# 1 Введение

Горение является одним из первых сложных физико-химических процессов, освоенных человечеством и использовавшихся им на протяжении многих тысячелетий. В настоящее время известно множество различных горючих смесей, например смесь метана $CH_4$, или окиси углерода CO с воздухом, или, так называемая, гремучая смесь водорода с кислородом. В таких смесях могут распространяться волны химической реакции [1].

Существуют два основных механизма распространения пламени в гомогенной газовой горючей смеси: теплопроводность и ударная волна. В первом случае, соседние с пламенем невозмущенные слои смеси нагреваются до температуры, при которой реакция протекает быстро, благодаря теплопроводностной передаче тепла от горящих слоев газа (продуктов горения). Скорость распространения пламени при этом не велика, так как теплопередача теплопроводностью – процесс сравнительно медленный. Процесс горения, движимый механизмом теплопроводности, называется медленным горением, или дефлаграцией.

Наряду с медленным горением существует и механизм быстрого распространения пламени – детонация. В детонационной волне воспламенение горючей смеси происходит в результате нагревания ее сильной ударной волной. Вслед за фронтом ударной волны начинается реакция горения смеси. Температура возрастает по сравнению с температурой за ударной волной вследствие выделения энергии и превращения химической энергии смеси в тепло. Зона реакции оказывается привязанной к ударной волне. Скорость детонационных волн в газе составляет 2–5 км/с, а в твердых и жидких ВВ типа тротила, нитроглицерина, гексогена – 6–8, даже до 9 км/с [2].

Фундаментальные исследования процессов горения, перехода горения в детонацию и самой детонации обычно проводятся в условиях, когда давление и температура горючей среды однородны по пространству. Такие наблюдения позволяют выявить основные характеристики взрывных процессов, изучить химико-кинетические свойства горючих смесей и прогнозировать развитие горения в условиях отсутствия локальных зон поджатия и разогрева энергоносителя. Однако в результате преднамеренного соблюдения однородности полей давления и температуры реализуется некоторый отрыв фундаментальных исследований от их практических приложений, например при описании выделения тепловой энергии в натуральных условиях [3].

При распространении взрывных волн в воздухе или при взаимодействии их с каким-нибудь препятствием происходят быстрые изменения давления, плотности, температуры и массовой скорости. Ударными волнами называются возмущения давления с конечным временем существования неизменного давления и постоянной температуры ударно-сжатого газа. Взрывными волнами считаются возмущения с непрерывным изменением давления и температуры за передним фронтом. Область взрывной волны, в которой давление превышает давление окружающей среды носит название «положительная фаза», а область, где давление меньше исходного, называется отрицательной фазой или фазой разрежения. Наибольший интерес представляют параметры взрывных волн, которые либо легко измерить, либо можно связать с определенным видом повреждения. Так, относительно просто измерить время до прихода ударной волны от центра взрыва в произвольную точку пространства и ее скорость, а также общее изменение давления от времени [4].

Интенсивность ударных и взрывных волн, входящих в сужающиеся каналы, значительно возрастает, что вследствие больших величин давлении и температуры может привести к детонационному возгоранию горючей смеси, присутствующей в области распространения волн [5]. Данный источник добавочной энергии в последнее время может рассматриваться в двух схожих ситуациях.

Во-первых, в двигателях внутреннего сгорания за счет искусственного подбора формы камер сгорания создаются структуры течения, обеспечивающие наивысшую эффективность воспламенения и последующего сжигания топлива. В камерах сгорания прямоточных воздушно-реактивных двигателей течение горючей газовой или двухфазной смеси, как правило, осложнено присутствием присоединенных и отраженных волн давления, формирующих сложную пространственно-временную структуру распределения температуры и давления [3, 11]. Во-вторых, самоподдерживающаяся детонационная волна может быть использована для метания снарядов, тел со скоростями 0.7–12 км/с [5].

В последнее время большой интерес вызывает задача детонации водорода в областях сложной геометрической формы. Примером являются исследования кумуляции волн с неплоской (сферической) формой фронта волны [5–11]. Так, в работе [6] выполнены систематические экспериментальные и расчетные исследования распространения ударных и детонационных волн в цилиндрических трубах и плоских каналах с двумя U-образными поворотами предельной кривизны. В качестве горючей смеси для исследования была выбрана стехиометрическая смесь пропана с воздухом. Численное решение уравнений Эйлера (первый этап) осуществлялось методом Годунова. Численное решение уравнений химической кинетики (второй этап) проводилось методом Рунге-Кутты четвертого порядка аппроксимации.

Метод Годунова для решений уравнений газовой динамики также был применен в работах [7, 8]. В первой из них были выполнены расчеты полей параметров в замкнутой кубической и полузамкнутых полостях при распространении в них ударно-волновых возмущений [7]. Во второй работе было разработано программное обеспечение, позволяющее в сжатые сроки проводить массовые параметрические расчеты, моделирующие нестационарные газодинамические течения; определен характер изменения степени фокусировки по давлению и температуре в зависимости от угла раскрытия конуса [8]. Под степенью фокусировки понимается отношение максимального значения параметра, полученного внутри объема в процессе счета, к значению, которое получилось бы при нормальном отражении от плоской стенки ударной волны с параметрами, равными локальным значениям их на входе в полость.

Исследование распространения взрывных и детонационных волн в конической области было проведено в работах [5, 8, 9]. В работе [5] получены зависимость давления в вершине конуса от времени, распределение давления по поверхности конуса и его оси после возгорания горючей смеси для разных моментов времени. Численное решение осуществлялось кончено-разностным двухшаговым методом Бурштейна, являющимся модификацией известного метода Лакса-Вендроффа. В работе же [9] система уравнений, описывающая конвективный перенос, решалась с помощью квазимонотонной консервативной схемы Годунова повышенного порядка точности [10]. Система уравнений, описывающая течение реагирующей газовой смеси, детальные кинетические механизмы горения водорода в воздухе и алгоритмы сопряжения газодинамики и химической кинетики приведены в [11]. Получены зависимости давления от времени в вершине конуса и точках, расположенных на его образующей. В данных расчетах варьировалась энергия взрыва (вес взрывчатого вещества) и концентрация водорода в смеси, при неизменной геометрии конуса.

В данной работе рассмотрена кумуляция ударных и детонационных волн в осесимметричной конической области. Расчеты проводятся при помощи центрально-разностной численной схемы второго порядка точности. Изучена кумуляция волн в конической области с характерным размером порядка 1 м. Получены картины течения в ударной и детонационной волнах в различные моменты времени, а также временные зависимости и максимальные значения давления при различных углах раскрытия конуса.

# 2 Постановка задачи

## 2.1 Уравнения газовой динамики

Распространение ударной волны внутри конуса, ее усиление за счет поперечных волн изучается в рамках нестационарных уравнений газовой динамики. Поскольку характерные времена исследуемых процессов очень малы (порядка сотен микросекунд), нет необходимости учитывать явления вязкости и теплопроводности. Поэтому в основу математической модели положены уравнения сохранения массы, количества движения и энергии для идеального многокомпонентного невязкого и нетеплопроводного газа (уравнения Эйлера), которые для цилиндрической системы координат можно записать в векторной форме:

$$\frac{\partial \mathbf{U}}{\partial t} + \frac{\partial \mathbf{F}}{\partial r} + \frac{\partial \mathbf{G}}{\partial z} + \frac{\mathbf{S}}{r} = 0, \qquad (2.1)$$

со следующими вектор-функциями:

$$\mathbf{U} = \begin{pmatrix} \rho \\ \rho u \\ \rho v \\ E \\ \rho Y_i \end{pmatrix}, \quad \mathbf{F} = \begin{pmatrix} \rho u \\ \rho u^2 + P \\ \rho uv \\ (E+P)u \\ \rho Y_i u \end{pmatrix}, \quad \mathbf{G} = \begin{pmatrix} \rho v \\ \rho uv \\ \rho v^2 + P \\ (E+P)v \\ \rho Y_i v \end{pmatrix}, \quad \mathbf{S} = \begin{pmatrix} 0 \\ 0 \\ 0 \\ Q \\ W_i \end{pmatrix}, (2.2)$$

где $t$ – время, $\mathbf{U}$ – вектор переменных, $\mathbf{F}$ и $\mathbf{G}$ – векторы потоков соответственно вдоль радиального и вертикального направлений, $\mathbf{S}$ – источниковый член, возникающий вследствие осевой симметрии области; $\rho$ – плотность газа, $u$ и $v$ – радиальная и вертикальная компоненты вектора скорости, $P$ – давление, $Y_i$ – массовая доля i-го компонента в смеси газов, $W_i$ – скорость образования (расходования) i-го компонента газовой смеси, $E$ – полная удельная энергия газа, представляющая сумму внутренней и кинетической энергий, $Q = \Delta H_c \cdot W_{H_2}$ – скорость энерговыделения за счет химических реакций, где $\Delta H_c = 120.9$ МДж/кг – удельная теплота сгорания водорода. Количество компонентов в газовой смеси – $N_s$. Для замыкания системы (2.1) используется уравнение состояния идеального газа:

$$P = \rho RT \sum_{i=1}^{N_s} Y_i \mu_i^{-1}, \quad e = \frac{1}{(\gamma-1)} \frac{P}{\rho}, \qquad (2.3)$$

где $\mu_i$ – молекулярный вес i-го компонента, а R – универсальная газовая постоянная.

Это приближение было использовано в [12, 13]; было показано [12], что при использовании уравнения состояния реального газа различие результатов было заметно лишь на начальном этапе. Непосредственно после взрыва и до конечного момента эта разница была очень мала.

Таким образом, при описании данной модели используется уравнение состояния идеального газа, показатель адиабаты полагается равным $\gamma = 1.4$; молекулярная масса воздуха $\mu = 29 \cdot 10^{-3} \frac{\text{кг}}{\text{моль}}$ (универсальная газовая постоянная $R = \frac{8.31}{\mu} = 287.6 \frac{\text{Дж}}{\text{кг} \cdot \text{К}}$).

Консервативные переменные в данной модели – $\rho$, $\rho u$, $\rho v$, $\rho Y_i$ и $E$. Остальные неизвестные (давление P, температура T, скорости u и v, удельная внутренняя энергия e, удельная кинетическая энергия $\frac{u^2+v^2}{2}$) вычисляются из консервативных переменных с применением соответствующих соотношений.

## *2.2 Кинетическая схема*

Горение водорода происходит по сложному цепному механизму, включающему образование и размножение активных радикалов. Для упрощения, в данной работе использована глобальная кинетическая схема Маринова [14], в которой рассматривается одна необратимая глобальная реакция горения водорода $H_2 + \frac{1}{2} O_2 = H_2O$. Константа скорости данной реакции при давлении 1 атм имеет вид:

$$k_{global} = 1.8 \times 10^{13} \exp(-17614 K/T) [H_2]^{1.0} [O_2]^{0.5}. \qquad (2.4)$$

## 2.3 Начальные и граничные условия

Использованы следующие начальные условия: при $t = 0$ $\rho = \rho_0$, $P = P_0$, $T = T_0$, $\rho u = \rho v = 0$, $E = \rho_0 e_0$, $Y_i = Y_{i0}$, где $\rho_0 = \frac{P_0}{RT_0}$, $e_0 = \frac{RT_0}{\gamma - 1}$ – плотность и внутренняя энергия невозмущенного газа. Состав газа $Y_{i0}$ соответствовал либо воздуху (при расчете ударных волн), либо смеси водорода с воздухом (при расчете детонации).

Ударные и детонационные волны инициировались путем выделения энергии Q в объеме $V_{в} = \frac{4}{3}\pi r_0^3$, имеющем сферическую форму радиуса $r = r_0$, с запасенной внутри энергией $E_{в} = E_0 + \frac{Q}{V_{в}}$. Предполагается, что энергия выделяется мгновенно, поэтому плотность и скорость не меняются во время взрыва, в то время как давление и температура возрастают.

На оси симметрии (левая граница расчетной области) установлены симметричные граничные условия для всех переменных, кроме радиального потока $\rho u$, который равен нулю вследствие осевой симметрии:

$$r = 0: \quad \rho u = 0, \quad \frac{\partial \varphi}{\partial r} = 0, \quad \varphi = \{\rho, \rho v, E, P, T, e, Y_i\}. \tag{2.5}$$

На верхней границе реакционного объема массовый поток $\rho v$ равен нулю и вертикальная составляющая градиента остальных неизвестных равна нулю:

$$z \in QSKLM: \quad \rho v = 0, \quad \frac{\partial \varphi}{\partial z} = 0, \quad \varphi = \{\rho, \rho u, E, P, T, e, Y_i\}. \tag{2.6}$$

На открытых (правой и нижней) границах расчетной области установлен нулевой градиент всех переменных, что позволяет газу покидать область через эти границы:

$$r = R_0: \quad \frac{\partial \varphi}{\partial r} = 0, \quad \varphi = \{\rho, \rho u, \rho v, E, P, T, e, Y_i\}, \tag{2.7}$$

$$z = 0: \quad \frac{\partial \varphi}{\partial z} = 0, \quad \varphi = \{\rho, \rho u, \rho v, E, P, T, e, Y_i\}. \tag{2.8}$$

# 3 Численный метод

## 3.1 Расчетная сетка

Решение задачи (2.1) – (2.3) осуществлялось численным интегрированием уравнений, входящих в данные соотношения. Внутри расчетной области $0 \leq r \leq R_0, 0 \leq z \leq Z_0$ введена равномерная сетка с $N_R \times N_Z$ ячейками, радиальный размер которых равен $\Delta r = R_0/N_R$, а вертикальный – $\Delta z = Z_0/N_Z$. Также для удобства аппроксимации граничных условий добавлен ряд фиктивных ячеек (снаружи расчетной области) вдоль каждой из границ, так что при расчете используется $(N_R + 2) \times (N_Z + 2)$ ячеек. Координаты границ ячеек вдоль радиального направления $r_b(i)$, $i = 1, \ldots, N + 1$ определяются во время генерации сетки, координаты же центра ячеек определяются из соотношения $r_c(i) = 0.5(r_b(i) + r_b(i-1))$, где $i = 2, \ldots, N_R + 1$ (аналогичным образом для вертикального направления). Размеры фиктивных ячеек совпадают с размерами внутренних. Расположение ячеек и их центров в расчетной области, а также их обозначение представлено на рис. 3.1. Пунктирными линиями показаны фиктивные ячейки. Все величины (давления, плотности, температуры, скорости, энергии и концентрации) определяются в центре ячеек.

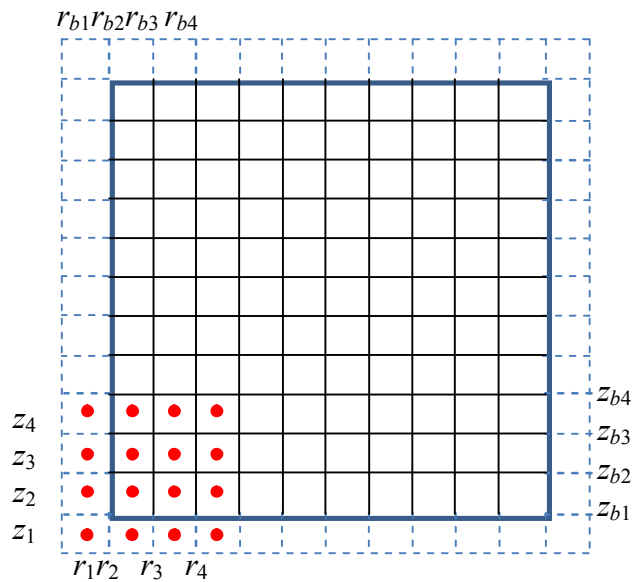

Рис. 3.1. Расположение и обозначение центров ячеек и их границ в расчетной области (фиктивные ячейки обозначены пунктирными линиями, точки соответствуют центрам ячеек, где рассчитываются все неизвестные).

### *3.2 Численная схема*

Рассмотрим аппроксимацию системы уравнений (2.1) – (2.3). Дискретизация основных уравнений выполняется с помощью монотонной центрально-разностной схемы высокого порядка [15]. Она обеспечивает большую точность по сравнению с простыми схемами первого порядка вследствие очень малой численной диффузии, вводимой в аппроксимацию.

Рассматриваемая численная схема [15] выводится для решения законов сохранения, записанных в форме

$$\frac{\partial}{\partial t}\mathbf{U}(r,z,t) + \frac{\partial}{\partial r}\mathbf{F}\big(\mathbf{U}(r,z,t)\big) + \frac{\partial}{\partial z}\mathbf{G}\big(\mathbf{U}(r,z,t)\big) + \frac{\mathbf{S}}{r} = 0, \qquad (3.1)$$

где $\mathbf{U}$ – вектор консервативных переменных длины N, и $\mathbf{F}$, $\mathbf{G}$ – нелинейные конвективные потоки соответственно вдоль радиального и вертикального направлений. Для реконструкции численных потоков на границе ячеек в схеме [15] используется монотонное центральное дифференцирование высокого порядка. Преимущество этой схемы по сравнению с противопоточными схемами в ее простоте, потому что не нужно определять собственные числа якобиана потока, что требуется в схемах Годунова.

В дискретизированной двумерной форме схема может быть переписана в виде

$$\frac{\mathbf{U}_{i,j}^{(n+1)} - \mathbf{U}_{i,j}^{(n)}}{\Delta t} + \frac{\mathbf{H}_{i+1/2,j} - \mathbf{H}_{i-1/2,j}}{\Delta r} + \frac{\mathbf{J}_{i,j+1/2} - \mathbf{J}_{i,j-1/2}}{\Delta z} = -\frac{1}{r}\mathbf{S}_{i,j}, \qquad (3.2)$$

где $\mathbf{H}$ и $\mathbf{J}$ – численные потоки на границах ячеек, рассчитываемые по формулам:

$$\mathbf{H}_{i+1/2,j} = \frac{1}{2}\big(\mathbf{F}(\mathbf{U}_{i+1/2,j}^{+}) + \mathbf{F}(\mathbf{U}_{i+1/2,j}^{-})\big) - \frac{a_{i+1/2,j}}{2}\big(\mathbf{U}_{i+1/2,j}^{+} - \mathbf{U}_{i+1/2,j}^{-}\big), \quad (3.3)$$

$$\mathbf{J}_{i,j+1/2} = \frac{1}{2}\big(\mathbf{G}(\mathbf{U}_{i,j+1/2}^{+}) + \mathbf{G}(\mathbf{U}_{i,j+1/2}^{-})\big) - \frac{a_{i,j+1/2}}{2}\big(\mathbf{U}_{i,j+1/2}^{+} - \mathbf{U}_{i,j+1/2}^{-}\big). \quad (3.4)$$

Промежуточные величины $\mathbf{U}_{i+1/2,j}^{\pm}$ в формулах (3.3) и (3.4) определяются из следующих соотношений

$$\mathbf{U}^+_{i+1/2,j} = \mathbf{U}_{i+1,j} - \frac{\Delta r}{2}(\mathbf{U}_r)_{i+1,j}, \quad \mathbf{U}^-_{i+1/2,j} = \mathbf{U}_{i,j} + \frac{\Delta r}{2}(\mathbf{U}_r)_{i,j}, \qquad (3.5)$$

где производные по пространству $(\mathbf{U}_r)_{i,j}$ аппроксимируются по значениям $(\mathbf{U}_r)_{i-1,j}$, $(\mathbf{U}_r)_{i,j}$ и $(\mathbf{U}_r)_{i+1,j}$ с помощью ограничителя minmod, который обеспечивает TVD свойство схемы:

$$(\mathbf{U}_r)_{i,j} = \min\operatorname{mod}\left(\theta\frac{\mathbf{U}_{i,j} - \mathbf{U}_{i-1,j}}{\Delta r}, \frac{\mathbf{U}_{i+1,j} - \mathbf{U}_{i-1,j}}{2\Delta r}, \theta\frac{\mathbf{U}_{i+1,j} - \mathbf{U}_{i,j}}{\Delta r}\right). \qquad (3.6)$$

Здесь $\operatorname{minmod}(a, b) = \frac{1}{2}[\operatorname{sign}(a) + \operatorname{sign}(b)] \cdot \min(|a|, |b|)$ и $1 \leq \theta \leq 2$.

Важно отметить, что в (3.6) каждая компонента вектора рассматривается отдельно, что упрощает схему, не ухудшая ее свойств. Аналогичные соотношения используются для z-направления. Стоит добавить, что устанавливая $(\mathbf{U}_r) = 0$ в (3.5), мы получаем схему с первым порядком точности, что в некоторых ситуациях может оказаться более устойчивым, чем использование схем с повышенным порядком.

Величины $a_{i+1/2,j}$ и $a_{i,j+1/2}$ в численных потоках (3.3) и (3.4) являются максимальными локальными скоростями распространения на границах ячеек, рассчитываемые из соотношений

$$a_{i+1/2,j} = \max\left\{\rho\left(\frac{\partial \mathbf{F}}{\partial \mathbf{U}}(\mathbf{U}^-_{i+1/2,j})\right), \rho\left(\frac{\partial \mathbf{F}}{\partial \mathbf{U}}(\mathbf{U}^+_{i+1/2,j})\right)\right\}, \qquad (3.7)$$

где $\rho(A) = \max_i|\lambda_i(A)|$ обозначает спектральный радиус матрицы A (максимальный модуль собственных чисел $\lambda_i(A)$). В случае уравнений газовой динамики (2.1) – (2.3) собственные числа известны: $u - c$, $u$ и $u + c$, где $c = (\partial P/\partial \rho)_s^{1/2}$ – локальная скорость звука, поэтому определение $a_{i+1/2,j}$ не требует решения задачи на нахождение собственных чисел.

Заметим, что для определения численных потоков (соотношения (3.3) – (3.7)) все значения берутся из предыдущего шага по времени. Условие на устойчивость и уменьшение полной вариации формулируются с использованием следующих чисел Куранта:

$$\text{CFL} = \frac{a\Delta t}{\Delta x} \leq \frac{1}{8}.\qquad(3.8)$$

*3.3 Учет сложной геометрии*

В настоящее время существует два основных способа учета сложной геометрии поверхности: использование криволинейных структурированных или неструктурированных расчетных сеток и метод поверхности уровня, в котором используется ортогональная декартова сетка, а наличие «погруженных» границ учитывается за счет корректировки аппроксимационных операторов. В настоящей работе используется метод поверхности уровня, который был разработан в 1980-х годах [16]. Данный метод легко учитывает изменение формы геометрии во времени, благодаря чему получил широкое применение в обработке изображений, компьютерной графике, вычислительной геометрии, вычислительной гидродинамике и др.

Рассмотрим систему уравнений (2.1). В случае декартовой системы координат данная система может быть переписана в виде:

$$\frac{\partial \mathbf{U}}{\partial t} + \nabla \cdot \mathbf{F} = 0,\qquad(3.9)$$

где $\mathbf{F}$ – вектор потоков. В случае, если граница $\Gamma(t)$ разделяет область $\Omega$ на две области $\Omega^1(t)$ и $\Omega^2(t)$, то, как и в задаче с несколькими жидкостями, описание изменения границы раздела проводится исходя из граничного условия, получаемого из двухкомпонентной задачи Римана:

$$\mathcal{R}(\mathbf{U}_{\text{fluid1}}, \mathbf{U}_{\text{fluid2}}) = 0 \text{ на } \Gamma(t).\qquad(3.10)$$

При рассмотрении задачи со сложной границей раздела, ее изменение определяется скоростью $\mathbf{v}_{\text{rg}}$. Тогда воздействие на границу определяется неполной задачей Римана:

$$\mathcal{R}(\mathbf{U}_{\text{fluid1}}, \mathbf{v}_{\text{rg}}) = 0 \text{ на } \Gamma(t).\qquad(3.11)$$

Рассмотрим газ, занимающий область $\Omega^1$ в двумерной декартовой системе координат с шагом по пространству $\Delta x$ и $\Delta y$. Конечно-объемная дискретизация может быть получена путем интегрирования системы (3.9) по всему пространственно-временному объему $\Delta_{ij} \cap \Omega^1(t)$ расчетной ячейки (i, j), занятой жидкостью

$$\int_n^{n+1} dt \int_{\Delta_{ij} \cap \Omega^1(t)} dx\, dy\, \frac{\partial \mathbf{U}}{\partial t}$$

$$+ \int_n^{n+1} dt \int_{\partial \Delta_{ij} \cap \Gamma(t)} dx\, dy\, \mathbf{F} \cdot \mathbf{n} = 0, \quad (3.12)$$

где $\partial \Delta_{ij}$ – все четыре грани ячейки, пересекающиеся с сеткой в точках $(x_i + \Delta x/2, y_j)$, $(x_i, y_j + \Delta y/2)$, $(x_i - \Delta x/2, y_j)$ и $(x_i, y_j - \Delta y/2)$. Обозначая расположение границы раздела Γ(t), $\partial \Delta_{ij} \cap \Gamma(t)$ может быть представлена в виде двух составляющих: первая представляет собой комбинацию четырех отрезков граней ячейки, которые пересекает граница раздела. Поэтому их можно записать в форме $A_{i+1/2,j}(t) \cdot \Delta y$, $A_{i,j+1/2}(t) \cdot \Delta x$, $A_{i-1/2,j}(t) \cdot \Delta y$ и $A_{i,j-1/2}(t) \cdot \Delta x$, где $1 \geq A \geq 0$ – апертура. Вторая же – отрезок границы раздела $\Delta \Gamma_{i,j}(t)$, лежащий внутри ячейки (i, j). Используя данные предположения, а также проводя явную аппроксимацию производных, получаем следующее соотношение [17]:

$$\alpha_{i,j}^{n+1} \mathbf{U}_{i,j}^{n+1} = \alpha_{i,j}^n \mathbf{U}_{i,j}^n + \frac{\Delta t}{\Delta x\, \Delta y} \widehat{\mathbf{X}}(\Delta \Gamma_{i,j}) + \frac{\Delta t}{\Delta x} \left[ A_{i-1/2,j} \widehat{\mathbf{F}}_{i-1/2,j} - A_{i+1/2,j} \widehat{\mathbf{F}}_{i+1/2,j} \right]$$

$$+ \frac{\Delta t}{\Delta y} \left[ A_{i,j-1/2} \widehat{\mathbf{F}}_{i,j-1/2} - A_{i,j+1/2} \widehat{\mathbf{F}}_{i,j+1/2} \right]. \quad (3.13)$$

где $\Delta t$ – шаг по времени, $\alpha_{i,j} \mathbf{U}_{i,j}$ – вектор консервативных переменных в разрезанной ячейке, $\widehat{\mathbf{F}}$ – усредненный поток через грань ячейки, $\widehat{\mathbf{X}}[\Gamma_{i,j}(t)]$ – среднее изменение момента и энергии через границу раздела, определяемое из соотношений (3.10) или (3.11). Следует отметить, что все величины в правой части получены на n-ом шаге по времени. Для целых ячеек, которых не пересекает граница раздела, объемные доли и апертуры равны единице, а соответствующий отрезок границы раздела $\Delta \Gamma_{i,j}(t)$ равен нулю. Это приводит к упрощению соотношения (3.13):

$$\mathbf{U}_{i,j}^{n+1} = \mathbf{U}_{i,j}^{n} + \frac{\Delta t}{\Delta x}(\hat{\mathbf{F}}_{i-1/2,j} - \hat{\mathbf{F}}_{i+1/2,j}) + \frac{\Delta t}{\Delta y}(\hat{\mathbf{F}}_{i,j-1/2} - \hat{\mathbf{F}}_{i,j+1/2}), \quad (3.14)$$

что соответствует обыкновенной конечно-объемной схеме, записанной в двумерной декартовой системе координат. С другой стороны, соотношение (3.13), применяемое на разрезанных ячейках, может рассматриваться как небольшая модификация (3.14) возле границы раздела.

Для описания положения границы раздела вводится функция $\varphi(x, y, t)$, равная расстоянию от текущей точки до границы, так что изоповерхность $\varphi(x, y, t) = 0$ соответствует границе раздела областей, причем $\nabla|\varphi| = 1$ [16, 17]. Зная $\varphi$, мы можем определить границу раздела $\Gamma(t) = \{x, y: \varphi(x, y, t) = 0\}$, которая разделяет всю область на две области, каждая из которых соответствует газу или твердому телу, в зависимости от знака функции $\varphi$. В данной работе апертура и объемная доля $A^+$ и $\alpha^+$ соответствуют области, занятой газом, где $\varphi > 0$, в то время как для области, недоступной газу ($\varphi < 0$), апертура и объемная доля обозначаются $A^-$ и $\alpha^-$.

В случае сложной геометрии поверхность уровня функции $\varphi(x, y, t) = 0$ определяется численно. Сначала нужно определить, лежит ли точка снаружи или внутри распространения течения. Для этого использован метод пересечения лучами [18, 19]. Затем, мы находим точку на границе раздела, имеющую минимальное расстояние до узла сетки, после чего обозначаем расстояние и направление на эту точку величиной $\varphi$ и ее нормальным направлением соответственно. На гране ячейки поверхность уровня функции $\varphi(x, y, t) = 0$ определяется из следующих соотношений:

$$\varphi_{i+1/2,j+1/2} = \frac{1}{4}(\varphi_{i,j} + \varphi_{i+1,j} + \varphi_{i,j+1} + \varphi_{i+1,j+1}),$$

$$\varphi_{i+1/2,j-1/2} = \frac{1}{4}(\varphi_{i,j} + \varphi_{i+1,j} + \varphi_{i,j-1} + \varphi_{i+1,j-1}),$$

$$\varphi_{i-1/2,j+1/2} = \frac{1}{4}(\varphi_{i,j} + \varphi_{i-1,j} + \varphi_{i,j+1} + \varphi_{i-1,j+1}), \quad (3.15)$$

$$\varphi_{i-1/2, j-1/2} = \frac{1}{4}(\varphi_{i,j} + \varphi_{i-1,j} + \varphi_{i,j-1} + \varphi_{i-1,j-1}).$$

Изменение знака функции φ вдоль расчетной ячейки означает, что данная ячейка пересечена границей раздела. Для таких ячеек, полагая линейное распределение φ, апертуры на гранях ячеек могут быть рассчитаны из соотношений:

$$A_{i+1/2,j} = a_{i+1/2,j} \text{ если } \varphi_{i+1/2,j+1/2} > 0 \quad \text{иначе } 1 - a_{i+1/2,j},$$

$$A_{i-1/2,j} = a_{i-1/2,j} \text{ если } \varphi_{i-1/2,j+1/2} > 0 \quad \text{иначе } 1 - a_{i-1/2,j},$$

$$A_{i,j+1/2} = b_{i,j+1/2} \text{ если } \varphi_{i-1/2,j+1/2} > 0 \quad \text{иначе } 1 - b_{i,j+1/2}, \qquad (3.16)$$

$$A_{i,j-1/2} = b_{i,j-1/2} \text{ если } \varphi_{i-1/2,j+1/2} > 0 \quad \text{иначе } 1 - b_{i,j-1/2},$$

где

$$a_{i\pm 1/2,j} = \frac{|\varphi_{i\pm 1/2,j+1/2}|}{|\varphi_{i\pm 1/2,j+1/2}| + |\varphi_{i\pm 1/2,j-1/2}|}, \quad b_{i,j\pm 1/2} = \frac{|\varphi_{i+1/2,j\pm 1/2}|}{|\varphi_{i+1/2,j\pm 1/2}| + |\varphi_{i-1/2,j\pm 1/2}|}. \qquad (3.17)$$

Если функция φ не меняет знак вдоль грани ячейки, то ее апертура равна либо 1 ($\varphi > 0$), либо 0 ($\varphi < 0$).

После определения поверхности уровня функции $\varphi(x, y, t) = 0$ объемные доли $\alpha_{i,j}^+$ в области, где $\varphi > 0$, могут быть аппроксимированы с помощью сглаженной функции Хевисайда $\alpha_{i,j}^+ = H(\varphi_{i,j}, \varepsilon)$, где ε – малый параметр порядка размеров расчетной ячейки. Объемные доли $\alpha_{i,j}^-$ в области, где $\varphi > 0$, могут быть рассчитаны из соотношения $\alpha^- = 1 - \alpha^+$. Сглаженная функция Хевисайда имеет следующий вид:

$$H(\varphi, \varepsilon) = \begin{cases} 0 & \varphi < -\varepsilon, \\ \frac{1}{2} + \frac{\varphi}{2\varepsilon} + \frac{1}{2\pi}\sin\left(\frac{\pi\varphi}{\varepsilon}\right) & -\varepsilon \leq \varphi \leq \varepsilon, \\ 1 & \varphi > \varepsilon, \end{cases} \qquad (3.18)$$

которая зависит от функции φ.

Для того чтобы получить изменение момента и энергии через границу раздела поверхностей, рассмотренные задачи Римана вместе с взаимодействием границы разрешаются на расчетных ячейках внутри области, расположенной рядом с границей. Для данных задач ищется решение вдоль нормали к границе раздела. Используя соотношение (3.11) для взаимодействия между газом и твердой границей, скорость границы раздела связана со скоростью по нормали следующим соотношением:

$$\mathbf{v}_I = (\mathbf{v}_{rg} \cdot \mathbf{N})\mathbf{N}, \qquad (3.19)$$

где $\mathbf{N} \equiv (N_x, N_y)$ – нормаль вдоль поверхности уровня. Следовательно, для области нахождения газа ($\varphi > 0$), перенос момента и энергии равен

$$\widehat{\mathbf{X}}^P(\Delta\Gamma) = p_I \Delta\Gamma \mathbf{N}_I \quad \text{и} \quad \widehat{X}^E(\Delta\Gamma) = p_I \Delta\Gamma \mathbf{N}_I \cdot \mathbf{v}_I, \qquad (3.20)$$

где $\Delta\Gamma$ и $\mathbf{N}_I$ длина (или площадь) и нормаль сегмента границы раздела, соответственно. Вектор $\widehat{\mathbf{X}}^P = (\widehat{X}_x^P, \widehat{X}_y^P)$ обозначает перенос момента вдоль осей x и y, в то время как $\widehat{X}^E$ отвечает за перенос энергии. В данной работе границы неподвижны ($\mathbf{v}_{rg} = 0$), следовательно, перенос энергии равен 0. Для переноса момента имеем следующие соотношения:

$$\widehat{X}_x^P(\Delta\Gamma_{i,j}) = p_I(A_{i+1/2,j} - A_{i-1/2,j}) \cdot \Delta x, \qquad (3.21)$$

$$\widehat{X}_y^P(\Delta\Gamma_{i,j}) = p_I(A_{i,j+1/2} - A_{i,j-1/2}) \cdot \Delta y. \qquad (3.22)$$

Как видно из рис. 3.2, длина отрезка границы раздела может быть найдена по формуле:

$$\Delta\Gamma_{i,j} = \sqrt{(A_{i+1/2,j} - A_{i-1/2,j})^2 + (A_{i,j+1/2} - A_{i,j-1/2})^2}. \qquad (3.23)$$

### *3.4 Расчет химической кинетики*

Для решения уравнений химической кинетики был использован пакет DVODE, разработанный в Ливерморской национальной лаборатории, который в случае нежесткой задачи определяет решение по формуле Адамса, а в случае жесткой задачи проводит решение методом Гира [20].

### *3.5 Численная реализация*

Численная схема, описанная в разделах 3.2, 3.3 реализована в программе DINGO (Detonation In Gases with Obstacles), написанной на языке FORTRAN 95. Программа снабжена графическим интерфейсом пользователя DingoView для ввода начальных данных, запуска расчетной программы DINGO и обработки результатов.

### *3.6 Тестовые расчеты: Одномерная задача Римана*

Рассмотрим одномерную систему уравнений Эйлера:

$$\frac{\partial}{\partial t}\begin{bmatrix}\rho\\m\\E\end{bmatrix} + \frac{\partial}{\partial x}\begin{bmatrix}m\\ \rho u^2 + p\\ u(E + p)\end{bmatrix} = 0, \quad p = (\gamma - 1)\cdot\left(E - \frac{\rho}{2}u^2\right), \quad (3.24)$$

где $\rho, u, m = \rho u, p$ и $E$ – плотность, скорость, момент, давление и полная энергия, соответственно. В данном примере вектор консервативных переменных $u = (\rho, m, E)^T$, а вектор потока равен $f(\vec{u}) = \big(m,\ \rho u^2 + p, u(E + p)\big)^T$. Используются следующие начальные данные:

$$\vec{u}(x, 0) = \begin{cases} u_L, & x < 0, \\ u_R, & x > 0. \end{cases} \quad (3.25)$$

Для проверки предложенной численной схемы и программы был проведен ряд тестов. Тестирование проводилось на широко известных задачах Сода и Лакса по распаду произвольного разрыва [21, 22]. В обеих задачах использовалось уравнение состояния идеального газа. В задаче Сода [21] начальные величины слева и справа от разрыва равны

$$u_L = (1, 0, 2.5)^T, \quad u_R = (0.125, 0, 0.25)^T.$$

Показатель адиабаты $\gamma$ в обеих задачах равен 1.4.

В результате распада разрыва, заданного начальными данными, образуются бегущие вправо ударная волна и контактный разрыв, а также волна разрежения, распространяющаяся влево от начального положения разрыва. На рис. 3.2 – 3.10 приведены графики плотности, давления и скорости для различного числа расчетных ячеек. Данная схема хорошо описывает ударную волну, однако немного размазывает контактный разрыв, что видно на рис. 3.5 – 3.6. В отличие от схем первого порядка данная схема хорошо воспроизводит волну разрежения. Отметим также незначительную немонотонность, которая появляется в области между контактным разрывом и ударной волной (рис. 3.5 – 3.6).

В задаче Лакса [22] начальные данные имеют вид

$$u_L = (0.445, 0.311, 8.928)^T, \qquad u_R = (0.5, 0, 1.4275)^T.$$

На рис. 3.11 – 3.19 приведены графики плотности, давления и скорости для различного числа расчетных ячеек. Видно, что данные профили хорошо согласуются с точным решением [22], однако, как и в случае решения задачи Сода, схема размазывает контактный разрыв. Таким образом, полученные результаты хорошо сходятся с известными в литературе численными решениями [21, 22].

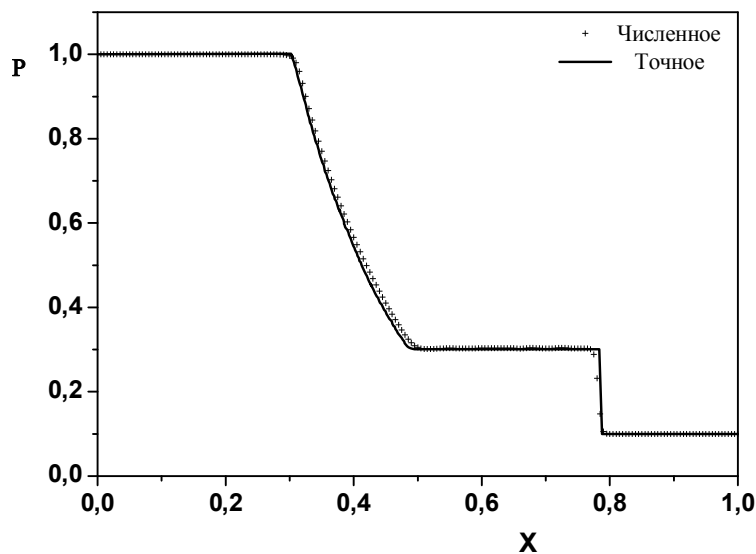

Рис. 3.2. Задача Сода – давление. $N = 200, t = 0.16$.

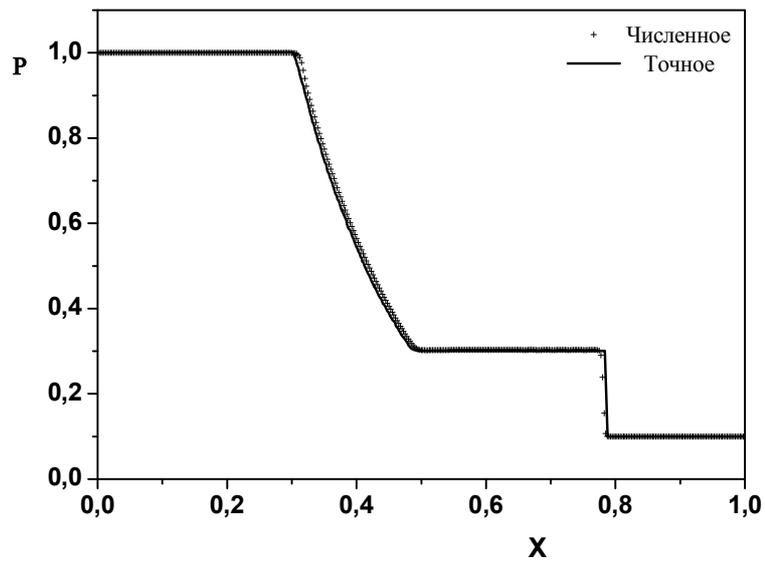

Рис. 3.3. Задача Сода – давление. $N = 400, t = 0.16$.

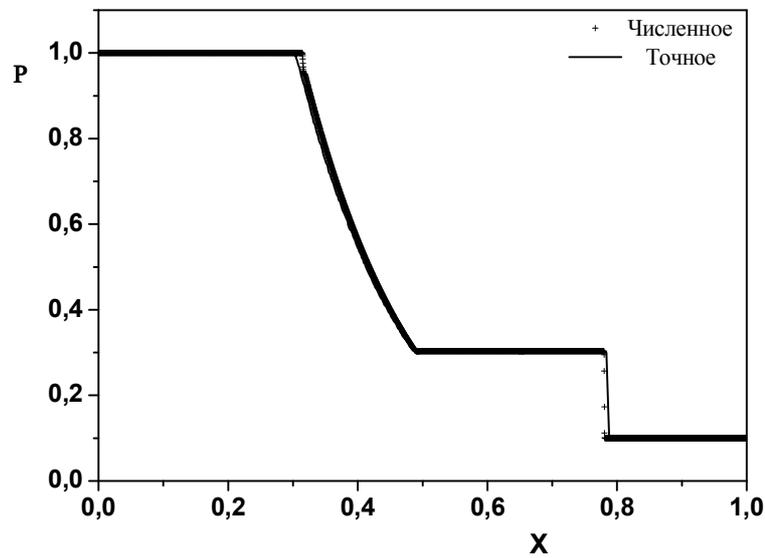

Рис. 3.4. Задача Сода – давление. $N = 4000, t = 0.16$.

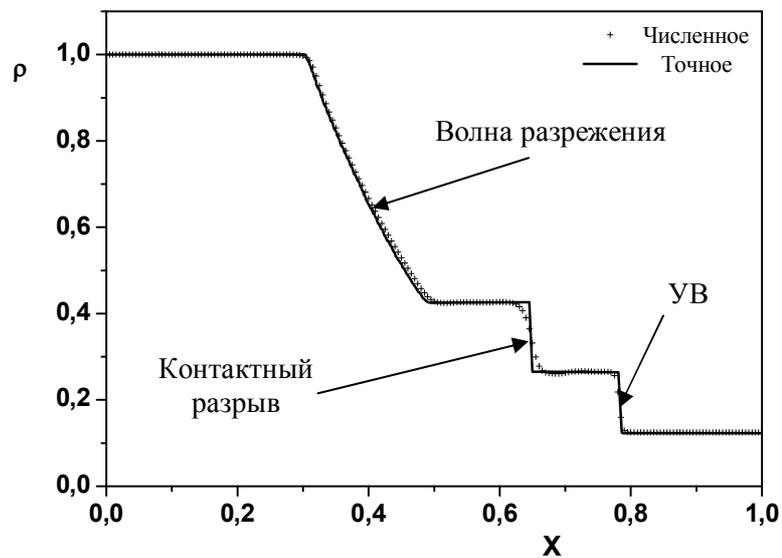

Рис. 3.5. Задача Сода – плотность. $N = 200, t = 0.16$.

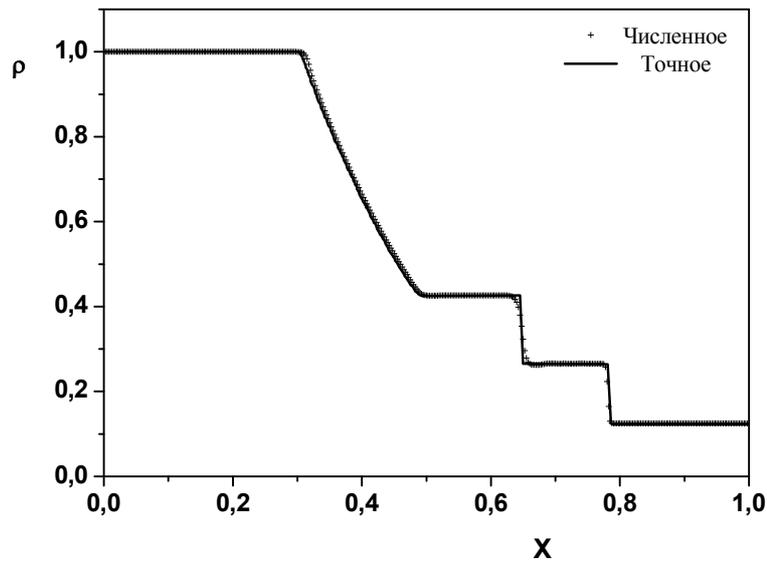

Рис. 3.6. Задача Сода – плотность. $N = 400, t = 0.16$.

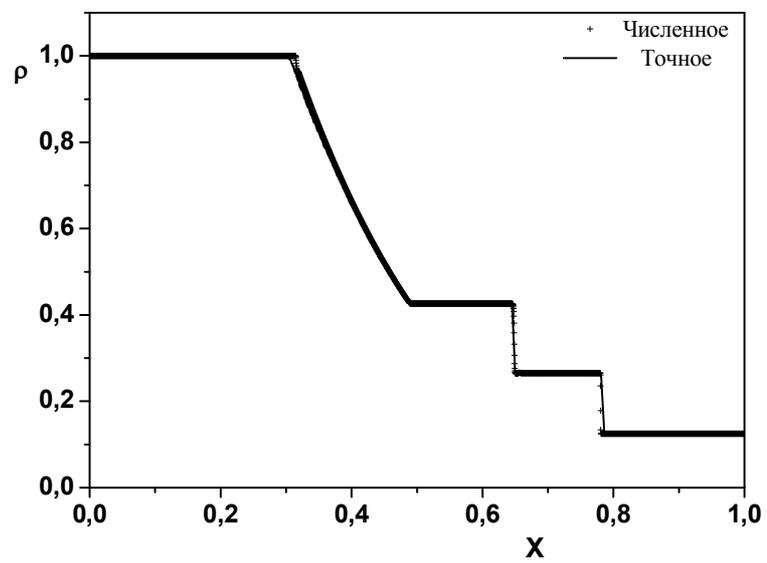

Рис. 3.7. Задача Сода – плотность. $N = 4000, T = 0.16$.

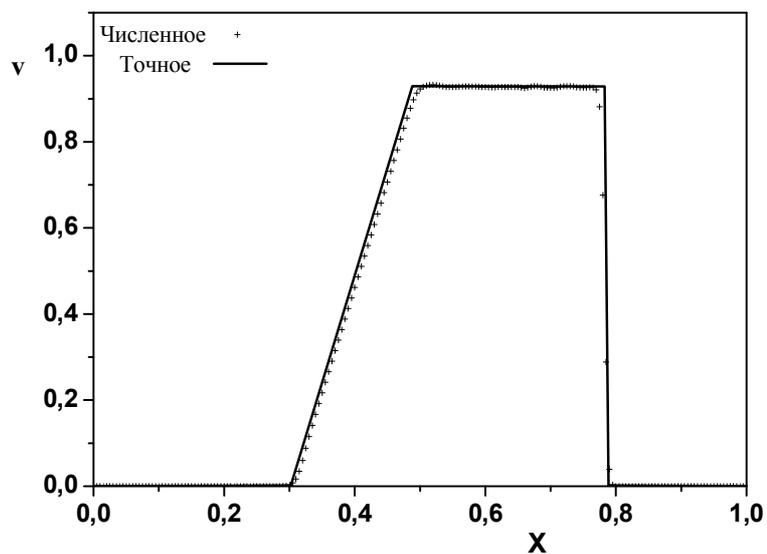

Рис. 3.8. Задача Сода – скорость. $N = 200, t = 0.16$.

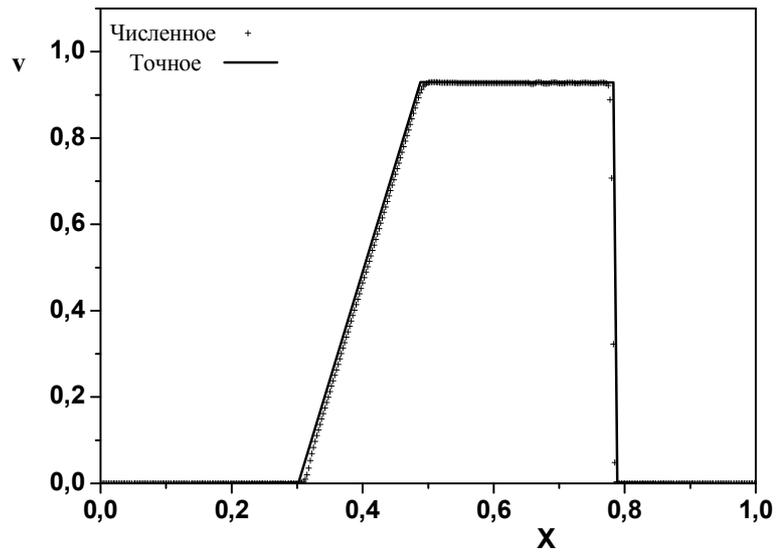

Рис. 3.9. Задача Сода – скорость. $N = 400, t = 0.16$.

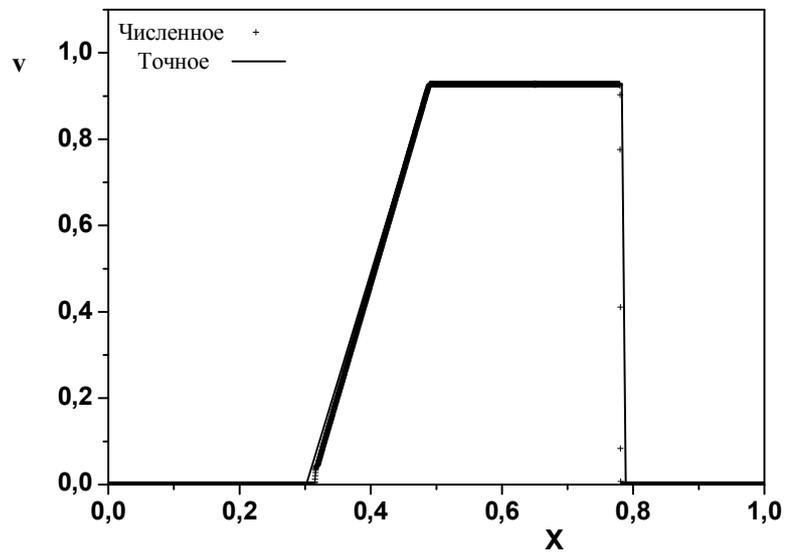

Рис. 3.10. Задача Сода – скорость. $N = 4000, t = 0.16$.

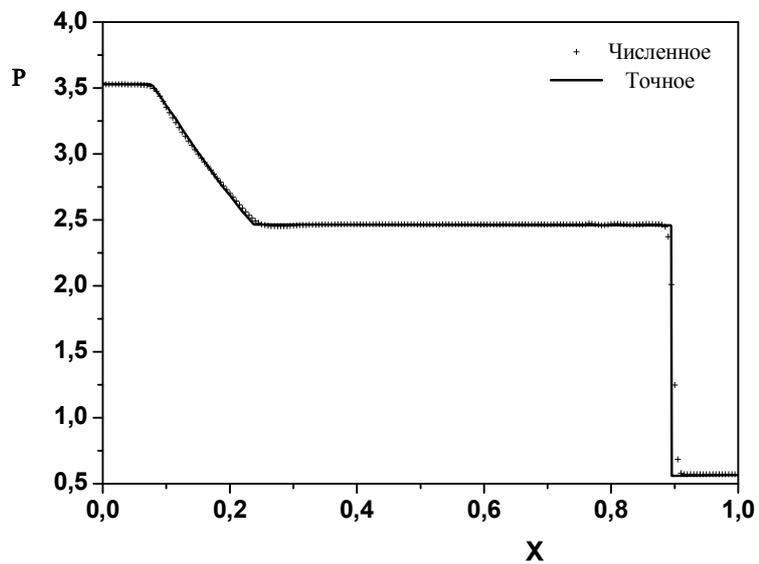

Рис. 3.11. Задача Лакса – давление. $N = 200, t = 0.16$.

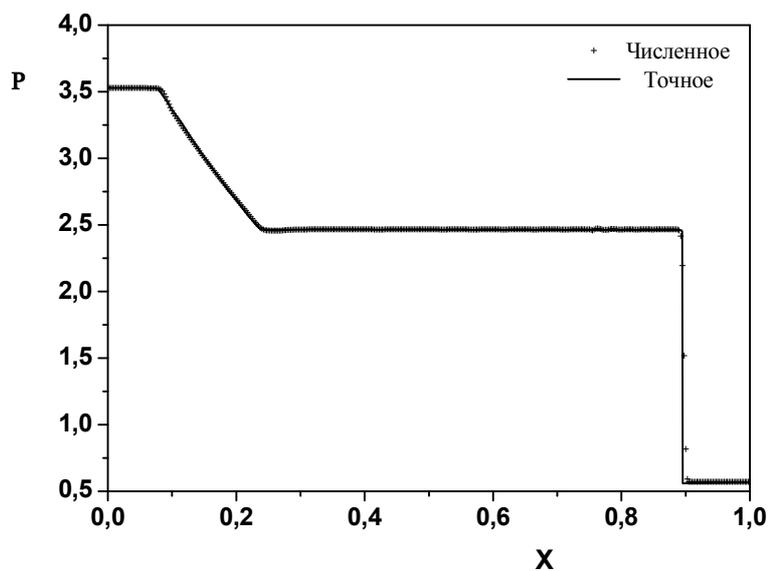

Рис. 3.12. Задача Лакса – давление. $N = 400, t = 0.16$.

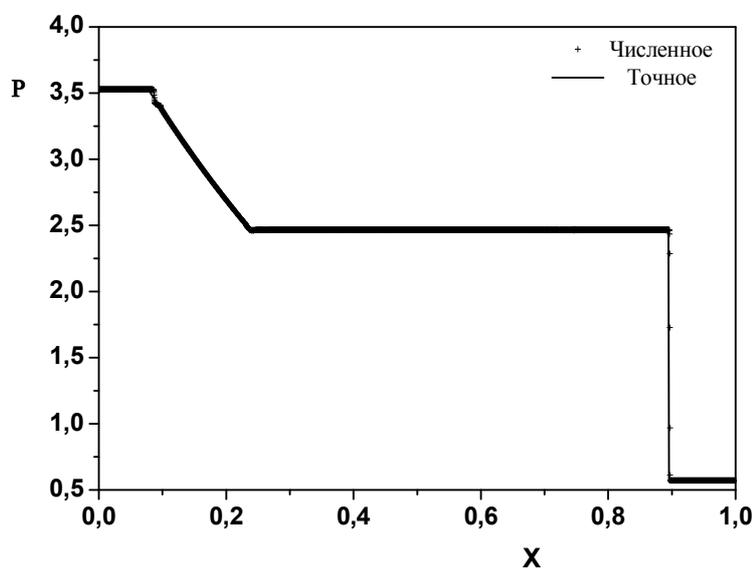

Рис. 3.13. Задача Лакса – давление. $N = 4000, t = 0.16$.

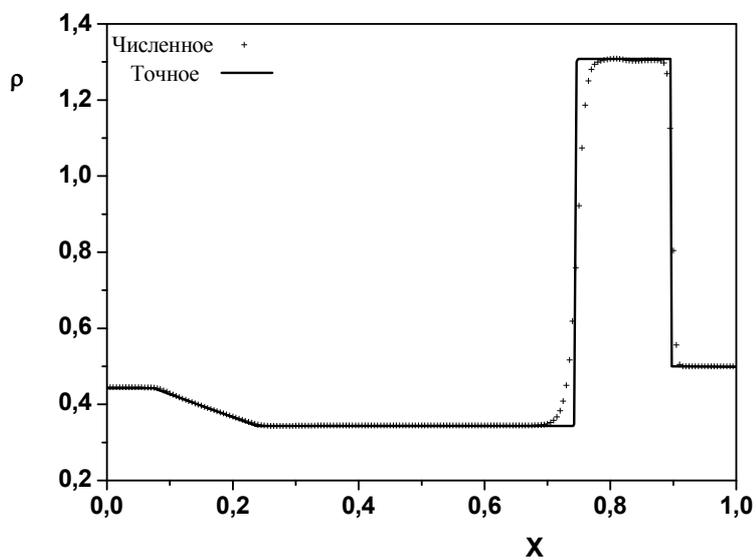

Рис. 3.14. Задача Лакса – плотность. $N = 200, t = 0.16$.

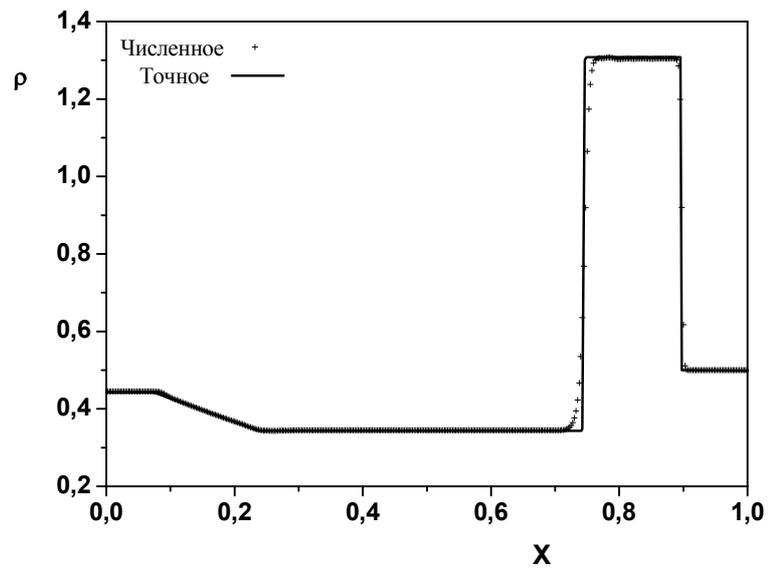
Рис. 3.15. Задача Лакса – плотность. $N = 400, t = 0.16$.

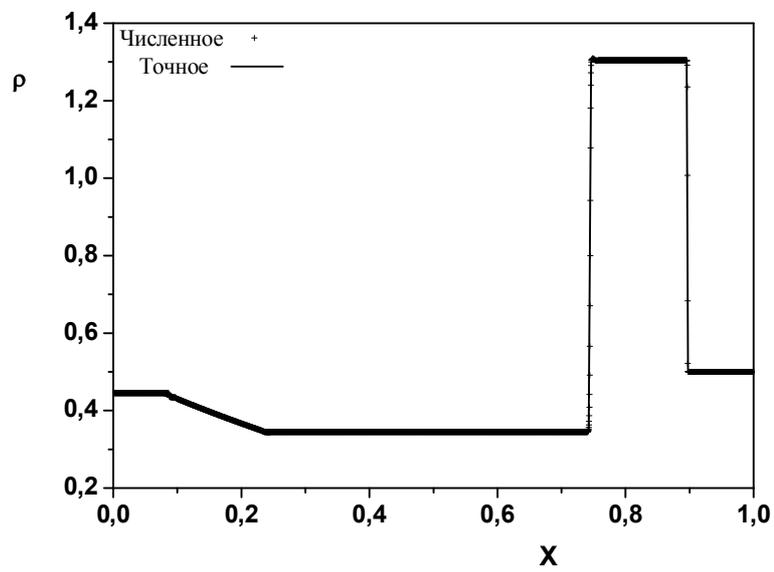
Рис. 3.16. Задача Лакса – плотность. $N = 4000, t = 0.16$.

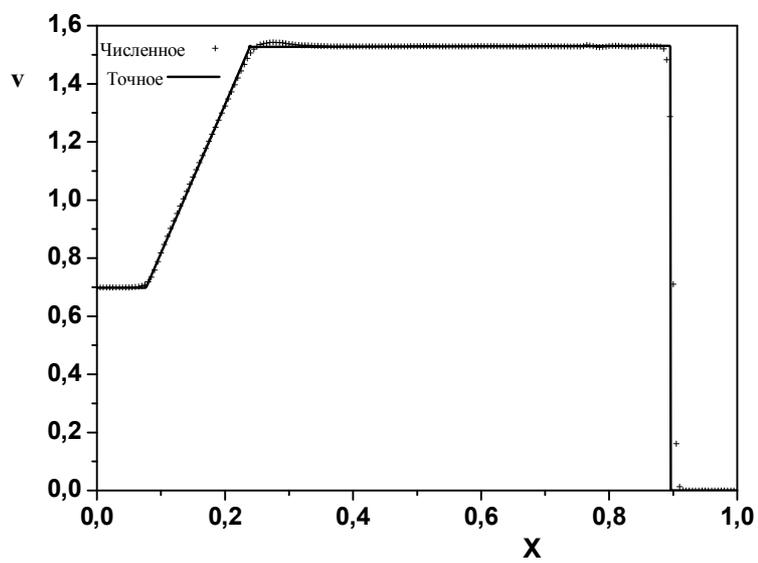
Рис. 3.17. Задача Лакса – скорость. $N = 200, t = 0.16$.

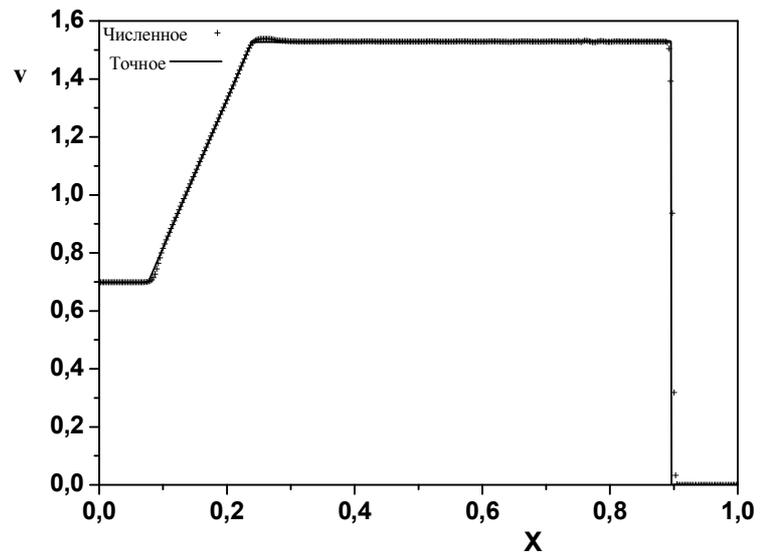

Рис. 3.18. Задача Лакса – скорость. N = 400, t = 0.16.

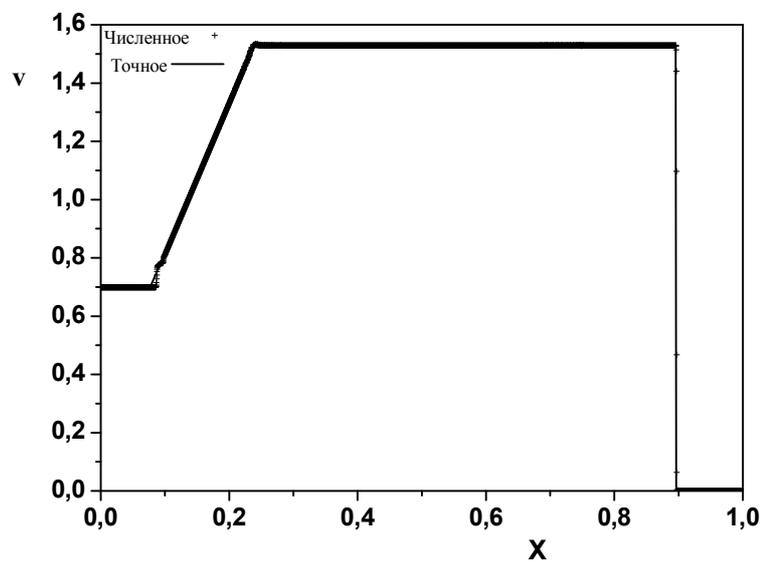

Рис. 3.19. Задача Лакса – скорость. N = 4000, t = 0.16.

# 4 Кумуляция ударных и детонационных волн в конической области

## *4.1 Геометрия*

Решение данной двумерной задач проводилось в цилиндрической области, в системе координат $(r, z)$. Инициирование производилось в источнике, расположенном на оси $z$, которая перпендикулярна горизонтальной поверхности.

Расчетная область представлена на Рис. 4.1. Рассматривается цилиндрическая область радиуса $R_0$ и высотой $Z_0$, левая граница которой совпадает с осью симметрии. Внутри расчетной области расположен реакционный объем, образованный конусом QSKL и границей LMNOQ. Острие конуса срезано так, что в вершине имеется малое затупление: угол при вершине равен 50°. Вдоль образующей конуса расположены датчики давления 1 – 5, на верхнем основании установлен датчик 6. Координаты вершин реакционного объема представлены в табл. 4.1.

Взрыв в источнике рассматривается как мгновенное выделение энергии в сферическом объеме радиуса $r_0$, центр которого находится на высоте $H_0$. Так как время выделения этой энергии мало, мы можем полагать, что плотность газа остается постоянной, в то время как его температура, а следовательно, давление возрастают. Таким образом, основная задача заключается в расчете дальнейшего распространения ударной волны и ее взаимодействия с геометрией.

Табл. 4.1. Координаты вершин реакционного объема.

| Вершина | r, мм | z, мм |
|---|---|---|
| K | 268 | 1114 |
| L | 268 | 1660 |
| M | 800 | 1660 |
| N | 800 | 0 |
| O | 0 | 0 |
| Q | 0 | 1660 |
| S | 13 | 1660 |

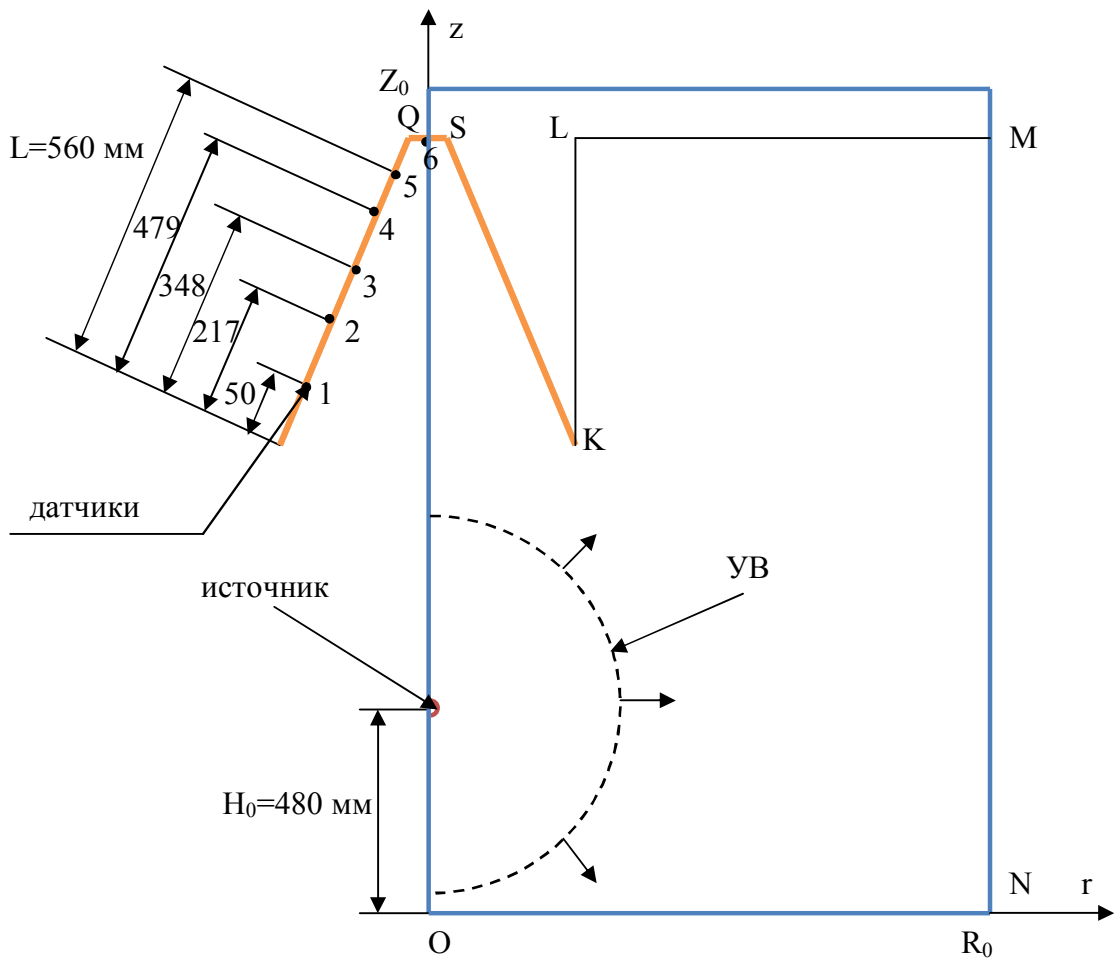

Рис. 4.1. Расчетная область.

## *4.2 Двумерные расчеты со сложной геометрией*

Были проведены численные расчеты распространения ударной волны в воздухе для двух вариантов расчетной области, представленных на рис. 4.2 и 4.3. Использовалась осесимметричная постановка задачи. Красным цветом показана область течения, синим – заблокированная область.

Помимо собственно расчетов кумуляции ударной волны, были проведены тестовые расчеты, направленные на проверку сходимости численных решений по отношению к

1) шагу сетки,

2) шагу по времени (числу Куранта),

3) взаимному влиянию «внешнего» и «внутреннего» течения, т. е. в конусе и вне его, с целью проверки реализации сложной геометрии и граничных условий.

Параметры расчетов представлены в табл. 4.2.

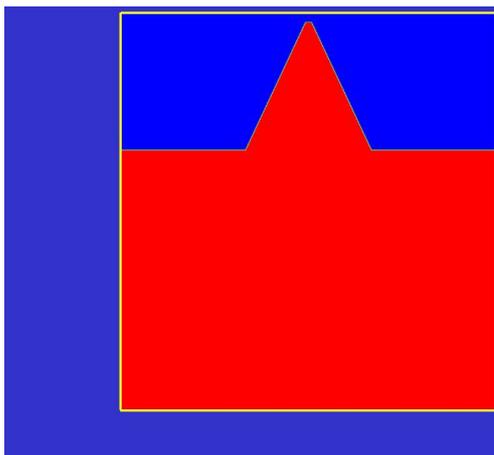

Рис.4.2. Расчетная область, вариант 1.

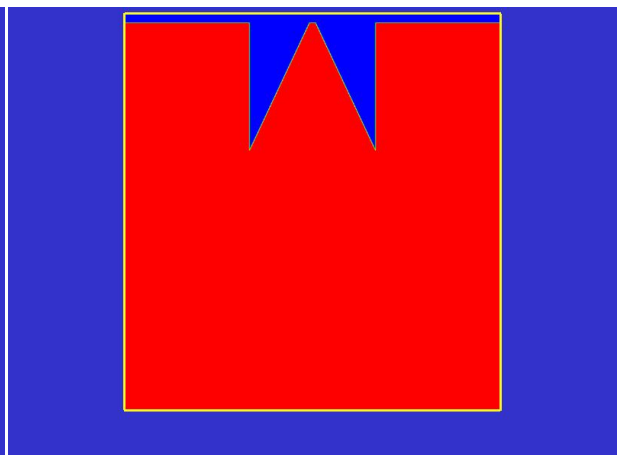

Рис.4.3. Расчетная область, вариант 2.

Табл. 4.2. Параметры численных расчетов.

| № расчета | Число Куранта | Сетка | Вариант геометрии | Инициирующий заряд, кДж |
|---|---|---|---|---|
| 1 | 0.125 | 200×400 | 1 | 15 |
| 2 | 0.125 | 200×400 | 2 | 15 |
| 3 | 0.125 | 400×800 | 1 | 15 |
| 4 | 0.125 | 400×800 | 2 | 15 |
| 5 | 0.25 | 200×400 | 2 | 15 |
| 6 | 0.0625 | 200×400 | 2 | 15 |

На рис. 4.4 – 4.32 представлены стадии кумуляции ударной волны в конусе (расчет 4). Показаны поля давления в различные моменты времени после выделения энергии в источнике (цветовая шкала приведена сверху, значения давления – в барах). Видно, что сферическая ударная волна достигает угла конуса к моменту времени t = 1200 мкс (см. рис. 4.9, 4.10). Взаимодействие ударной волны с внутренней поверхностью конуса приводит к ее отражению и формированию ударной волны, сходящейся к вершине конуса (рис. 4.11 – 4.16). В момент t = 2000 мкс отраженные волны с противоположных сторон конуса «схлопываются» (рис. 4.17). Отражение от донышка конуса происходит в момент t = 2600 мкс (рис. 4.23).

Далее, отраженная ударная волна движется наружу, причем в верхней части конуса формируется область низкого давления (рис. 4.26 – 4.32).

На рис. 4.33 – 4.38 приведены показания датчиков давления 1 – 6

соответственно. На каждом графике построены зависимости давления от времени для всех расчетов, представленных в таблице 4.2. Сравнение кривых позволяет оценить влияние расчетной сетки и шага по времени (числа Куранта) на полученные решения. При уменьшении пространственного шага расчётной сетки по обеим координатам (r, z) в два раза наблюдается некоторое смещение максимумов давления и изменение амплитуды волны. Величина этого смещения составляет не более 10% от показания датчика давления (рис. 4.33 – 4.38). При изменении же числа Куранта в 2 раза относительное изменение давления не превышает 1%.

Кроме того, зависимости давления от времени, полученные при различной геометрии конуса (см. рис. 4.2 и 4.3), отличаются не более чем на 1%. Также на рис. 4.39 – 4.44 приведены показания датчиков плотности 1 – 6 соответственно, а на рис. 4.45 – 4.50 приведены показания датчиков температуры 1 – 6. Качественные выводы совпадают со сделанными выше для датчиков давления.

На рис. 4.51 представлены зависимости максимального и минимального давления в расчетной области от времени, полученные для расчетов 2 и 4 (т.е. на более грубой и более мелкой сетках). Видно, что до момента начала взаимодействия ударной волны с конусом (t = 1300 мкс) максимальные значения практически совпадают. Это свидетельствует о малом влиянии схемной вязкости для применяемой численной схемы высокого порядка.

Наибольшие отличия в максимальных давлениях наблюдаются на интервале времени между t = 2000 и 2500 мкс, когда происходит «схлопывание» отраженных от боковой поверхности конуса ударных волн. Из рис. 4.17 – 4.23 видно, что при схлопывании волн возникает острый пик давления на оси конуса. Величина максимума оказывается чувствительной к разрешению расчетной сетки. Однако значения максимального давления при отражении от донышка конуса (t = 2600 мкс), полученные на обеих сетках, практически совпадают и составляют 5.5 атм. Кроме того, получено хорошее совпадение минимальных давлений.

Таким образом, тестовые расчеты показывают, что схема и ее реализация позволяют производить численные расчеты ударной волны в области сложной геометрии.

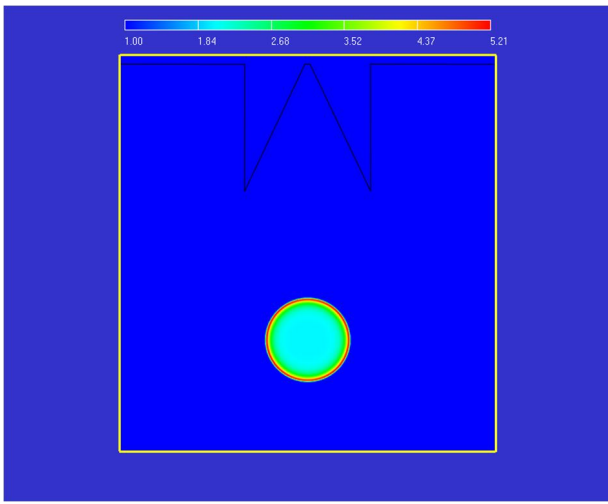

Рис. 4.4. Поле давления при кумуляции УВ в момент t = 100 мкс.

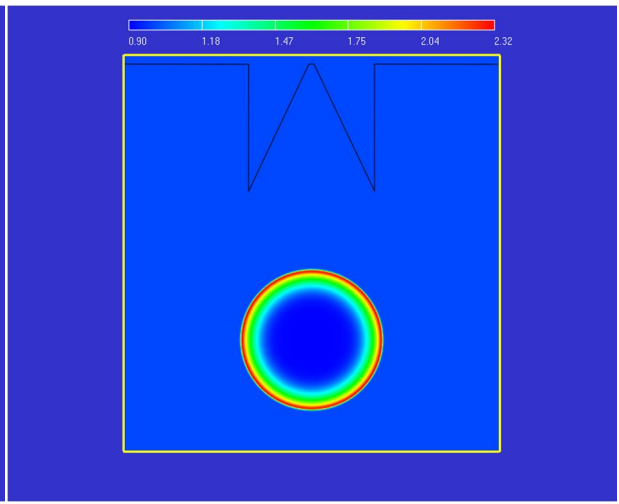

Рис. 4.5. Давление в УВ (t = 300 мкс).

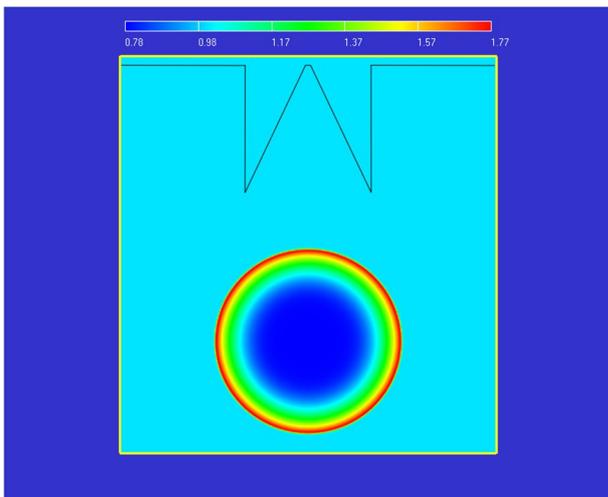

Рис. 4.6. Давление в УВ (t = 500 мкс).

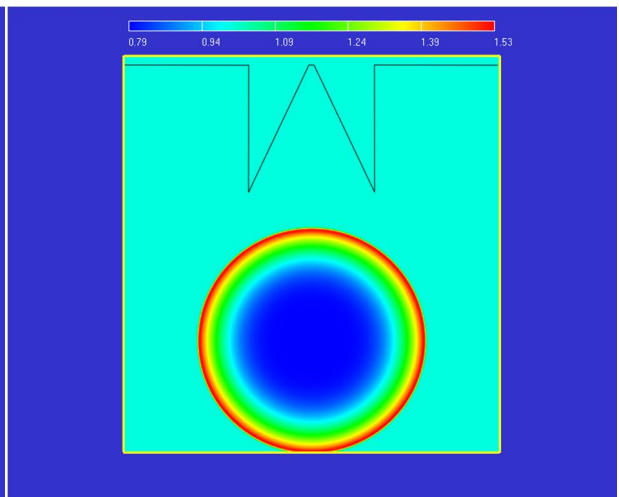

Рис. 4.7. Давление в УВ (t = 700 мкс).

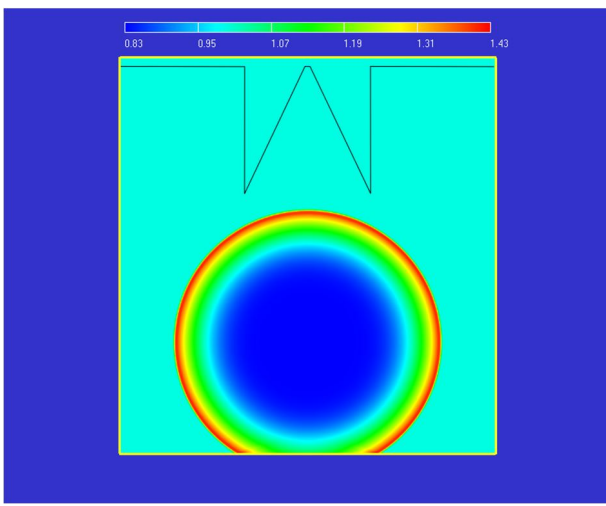

Рис. 4.8. Давление в УВ
(t = 900 мкс).

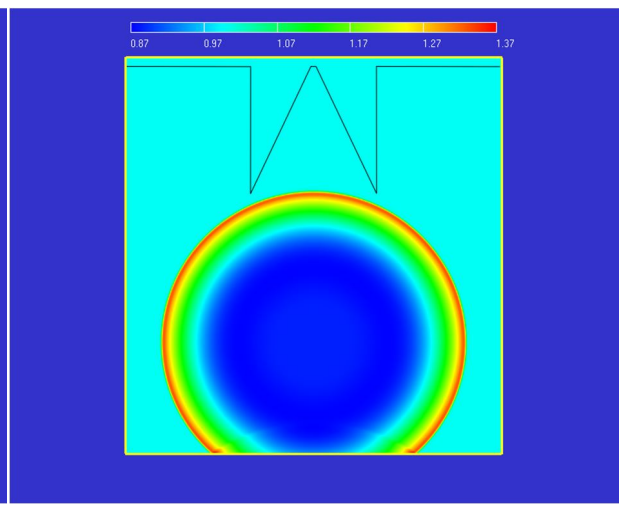

Рис. 4.9. Давление в УВ
(t = 1100 мкс).

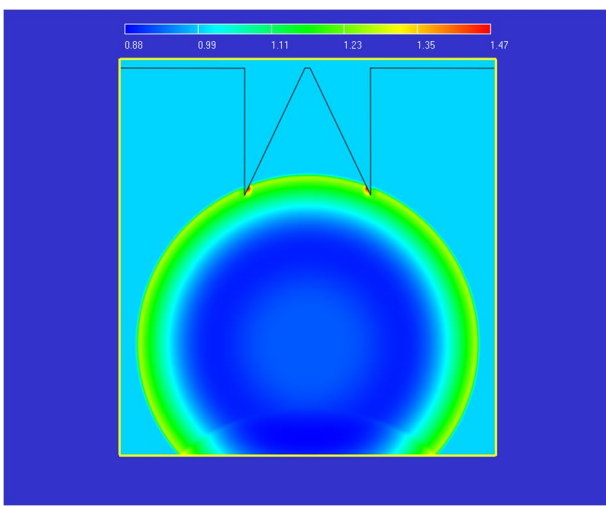

Рис. 4.10. Давление в УВ
(t = 1300 мкс).

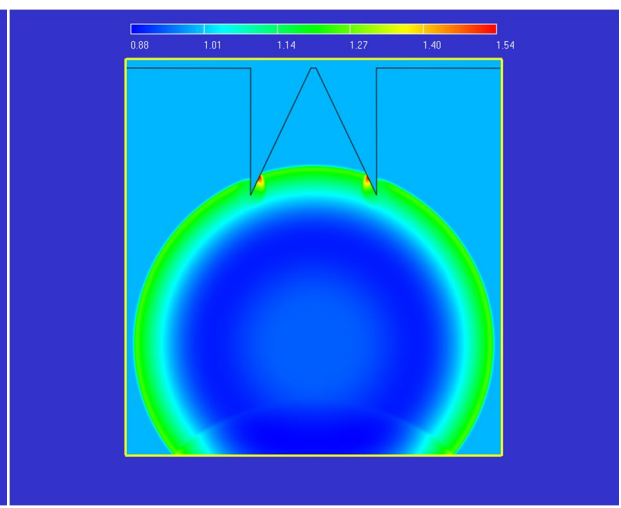

Рис. 4.11. Давление в УВ
(t = 1400 мкс).

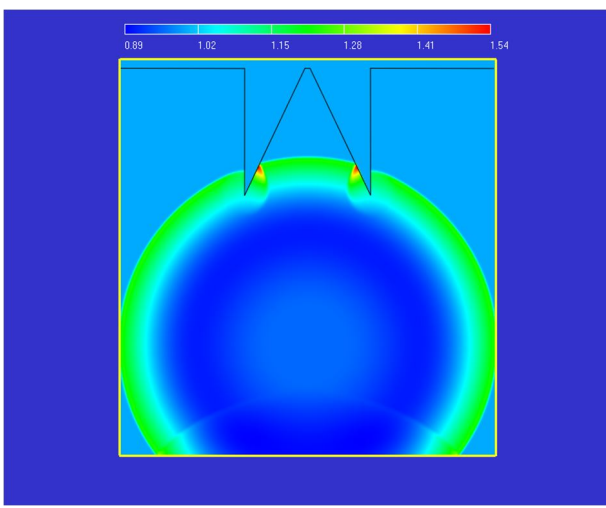

Рис. 4.12. Давление в УВ
(t = 1500 мкс).

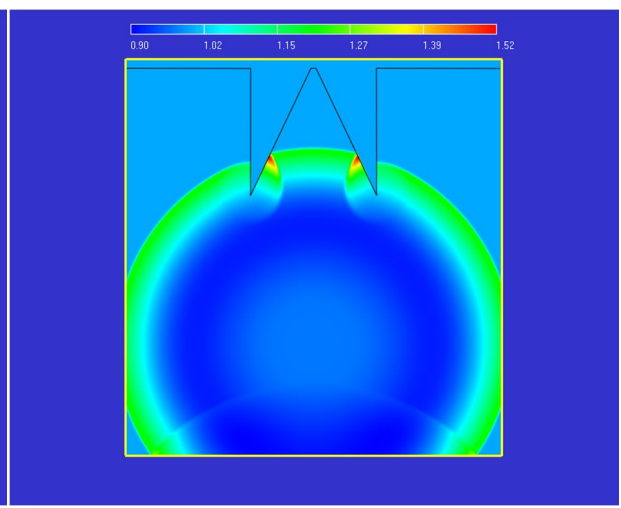

Рис. 4.13. Давление в УВ
(t = 1600 мкс).

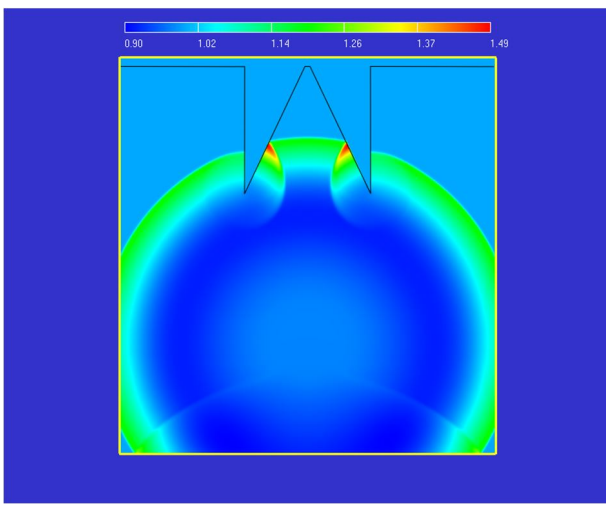
Рис. 4.14. Давление в УВ
(t = 1700 мкс).

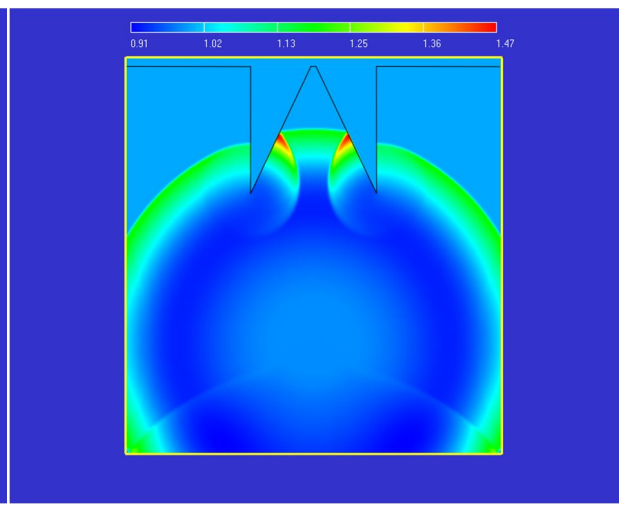
Рис. 4.15. Давление в УВ
(t = 1800 мкс).

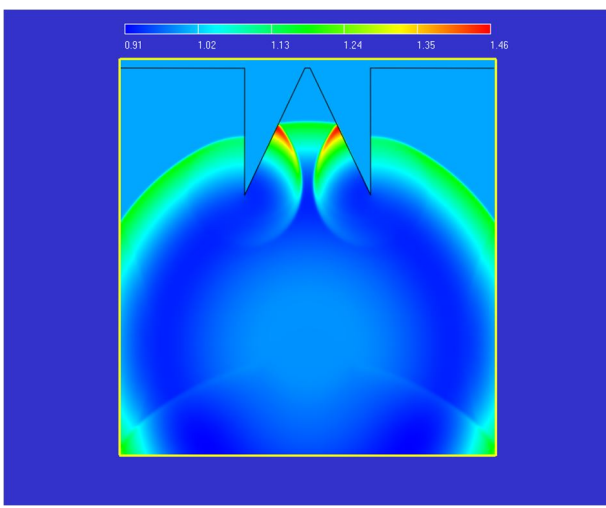
Рис. 4.16. Давление в УВ
(t = 1900 мкс).

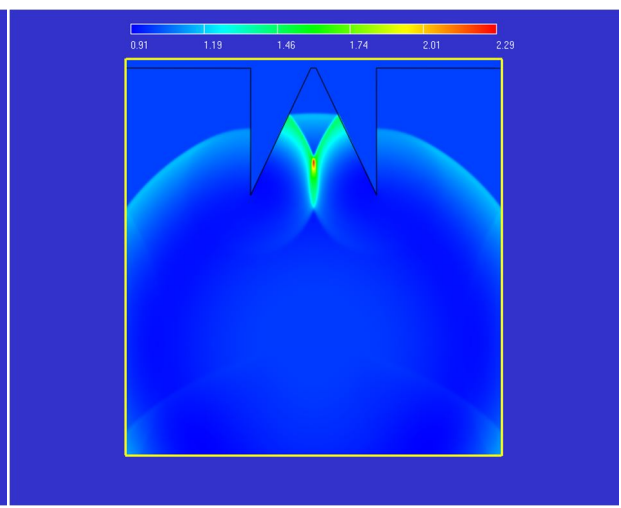
Рис. 4.17. Давление в УВ
(t = 2000 мкс).

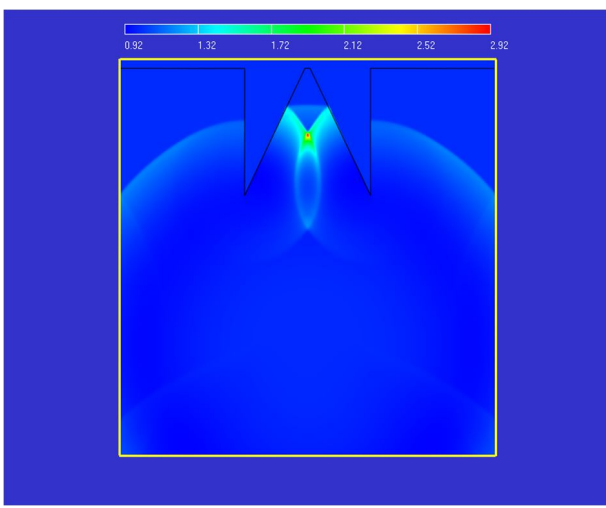
Рис. 4.18. Давление в УВ
(t = 2100 мкс).

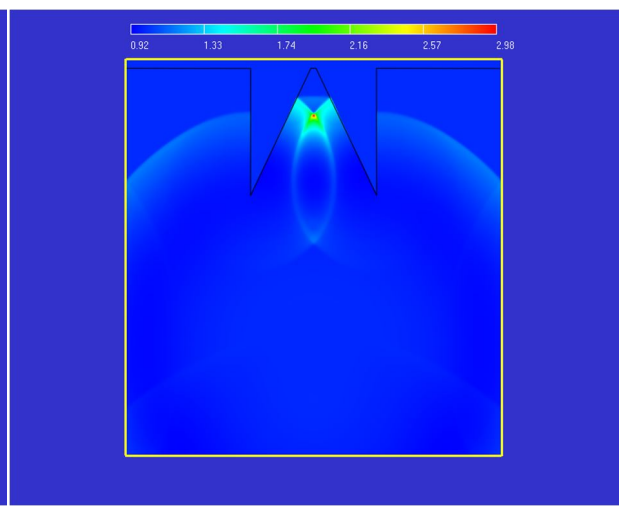
Рис. 4.19. Давление в УВ
(t = 2200 мкс).

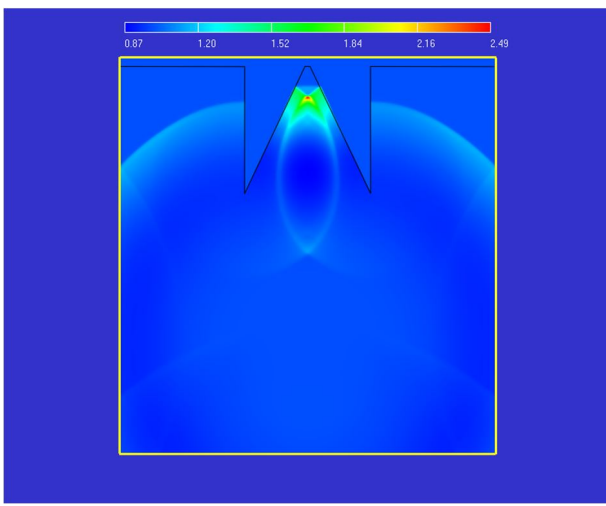
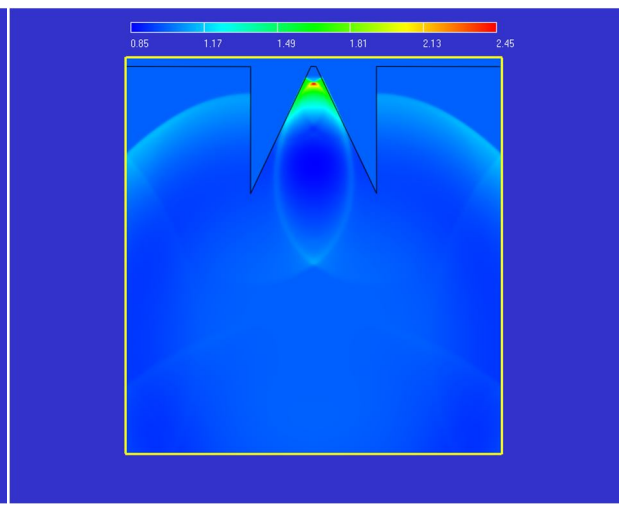

Рис. 4.20. Давление в УВ (t = 2300 мкс).

Рис. 4.21. Давление в УВ (t = 2400 мкс).

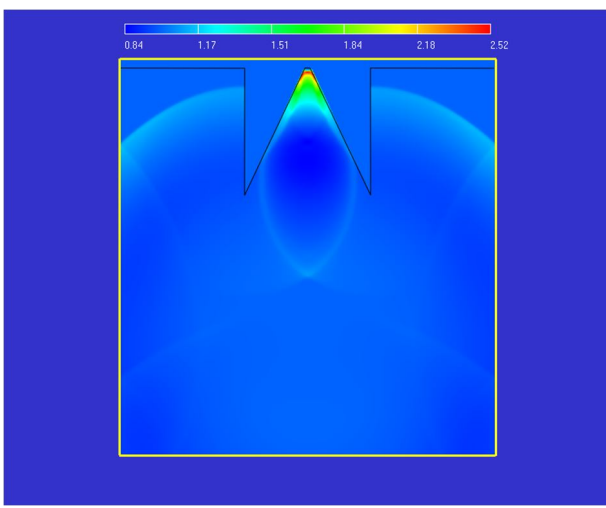
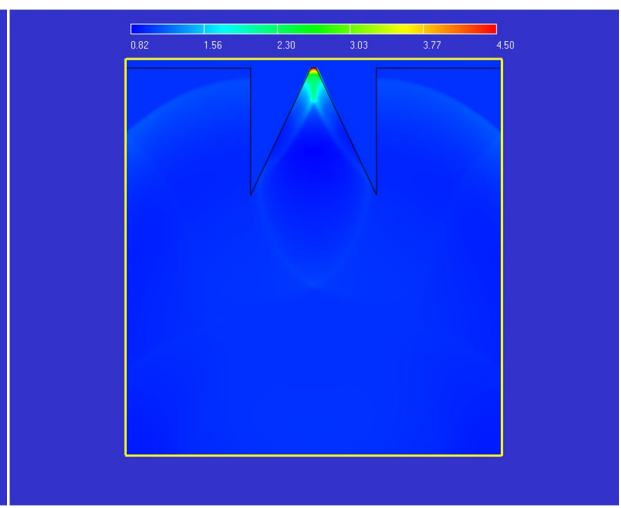

Рис. 4.22. Давление в УВ (t = 2500 мкс).

Рис. 4.23. Давление в УВ (t = 2600 мкс).

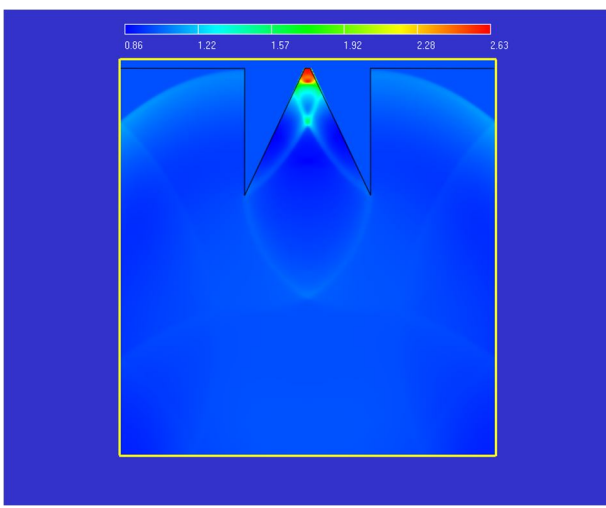
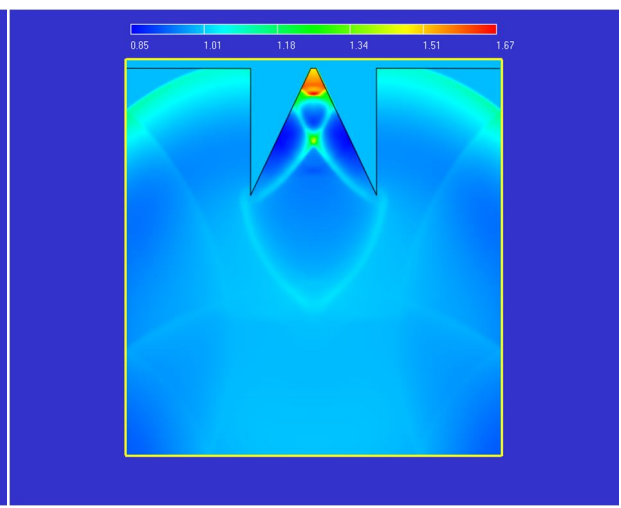

Рис. 4.24. Давление в УВ (t = 2700 мкс).

Рис. 4.25. Давление в УВ (t = 2800 мкс).

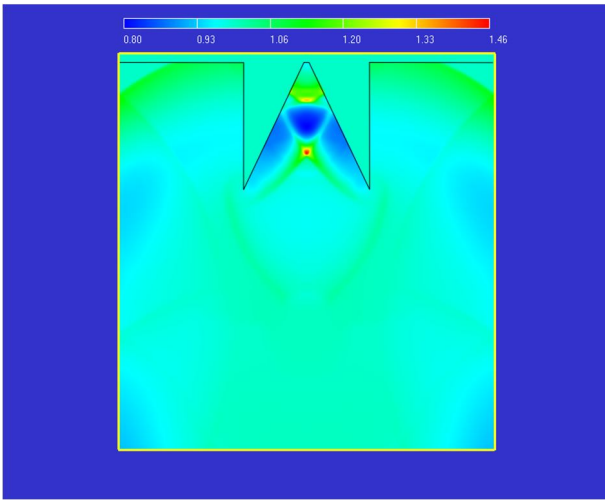

Рис. 4.26. Давление в УВ
(t = 2900 мкс).

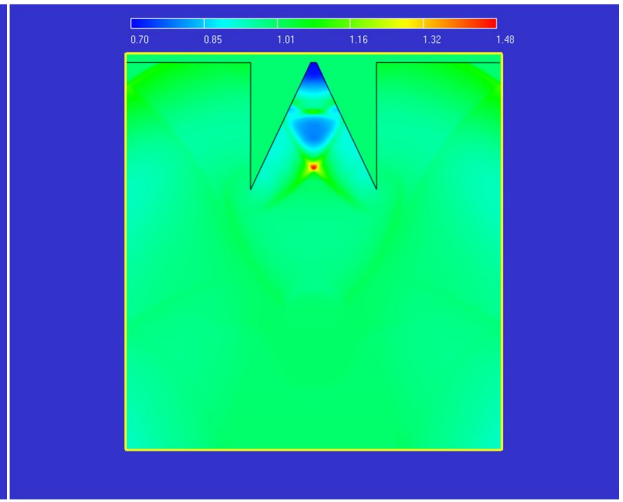

Рис. 4.27. Давление в УВ
(t = 3000 мкс).

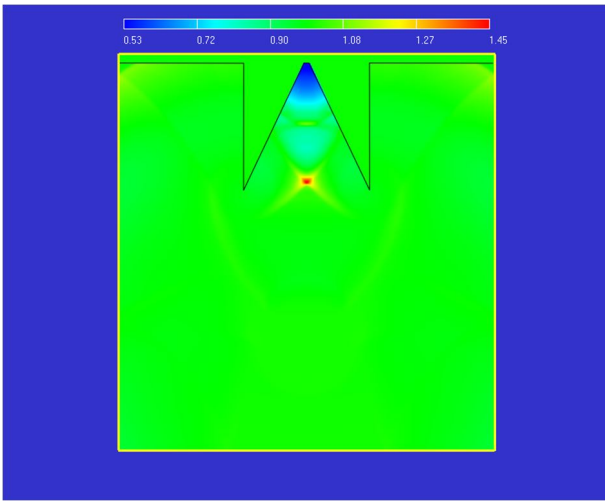

Рис. 4.28. Давление в УВ
(t = 3100 мкс).

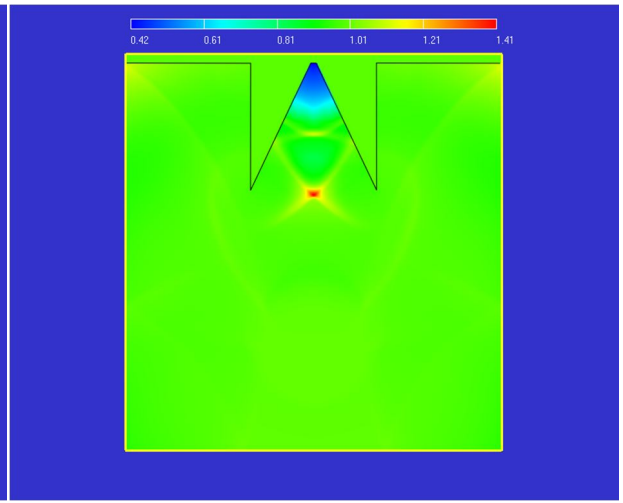

Рис. 4.29. Давление в УВ
(t = 3200 мкс).

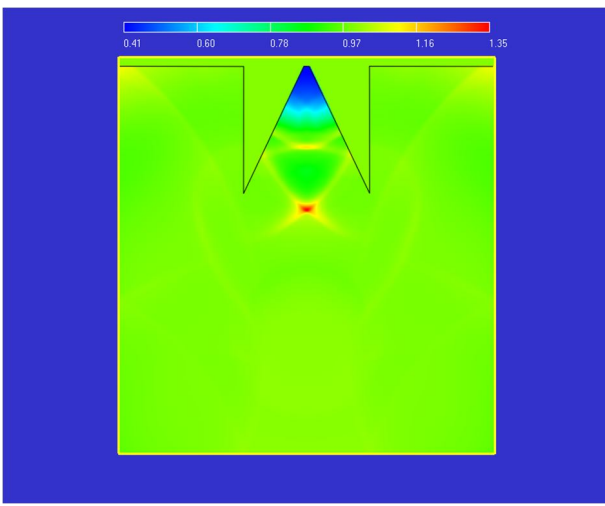

Рис. 4.30. Давление в УВ
(t = 3300 мкс).

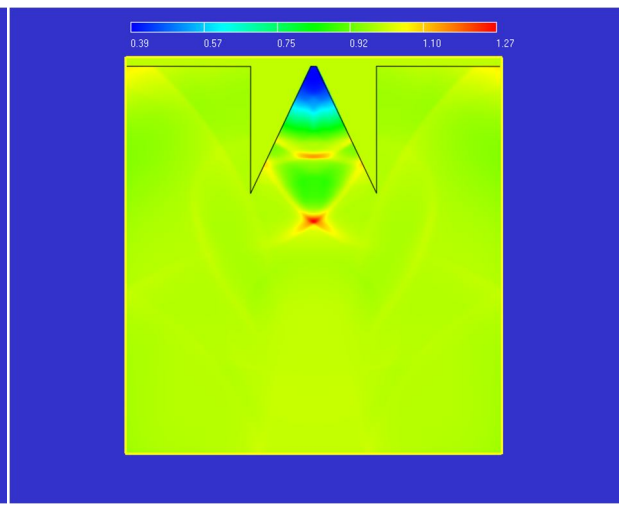

Рис. 4.31. Давление в УВ
(t = 3400 мкс).

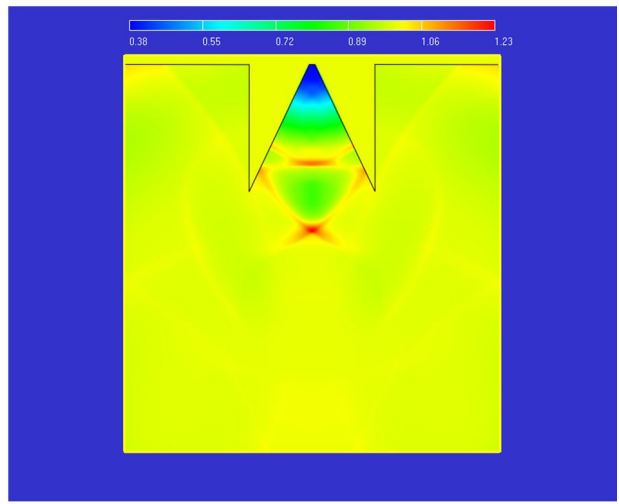

Рис. 4.32. Давление в УВ (t = 3500 мкс).

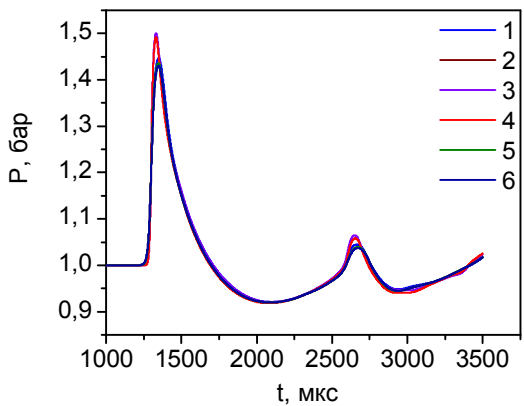

Рис. 4.33. Показания датчика давления 1.

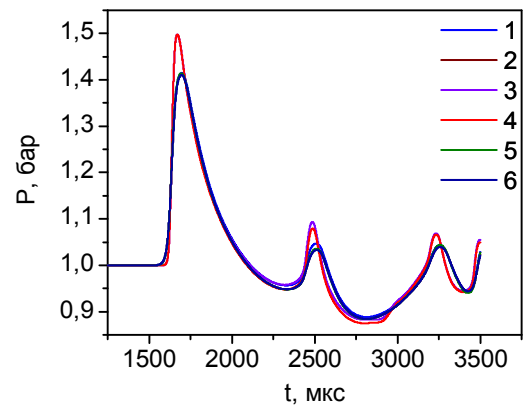

Рис. 4.34. Показания датчика давления 2.

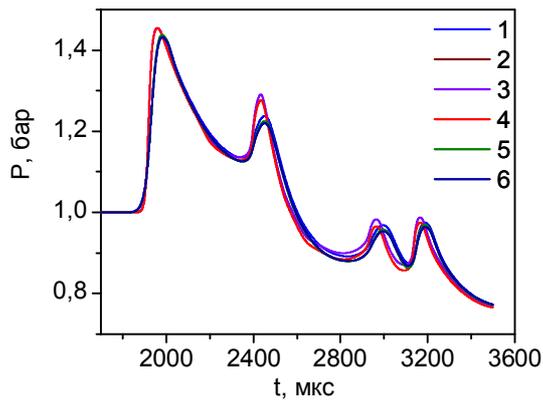

Рис. 4.35. Показания датчика давления 3.

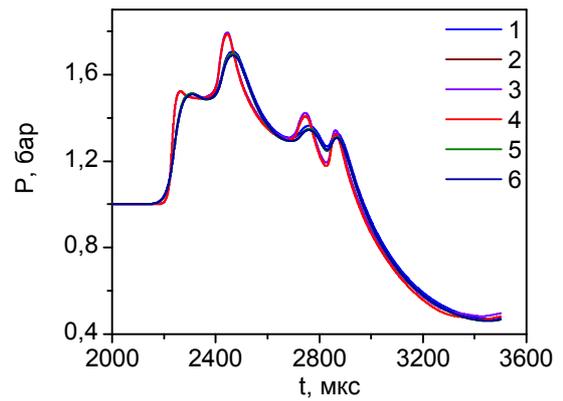

Рис. 4.36. Показания датчика давления 4.

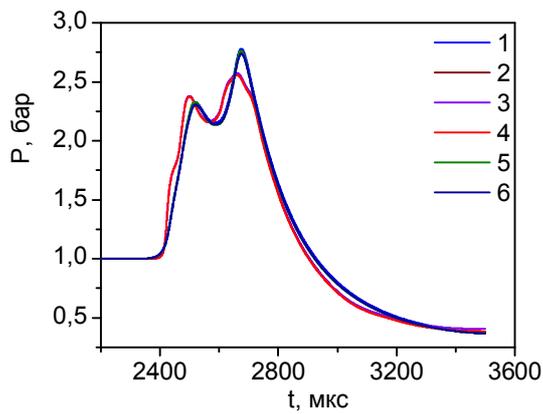

Рис. 4.37. Показания датчика давления 5.

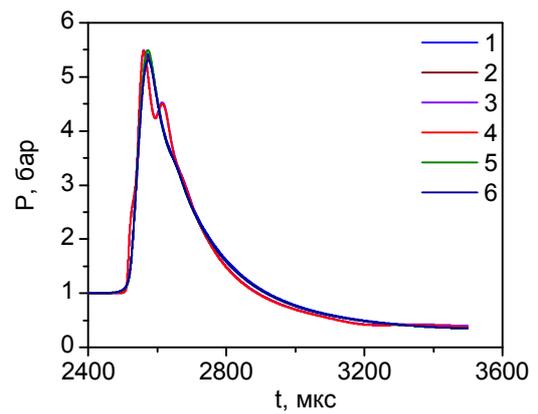

Рис. 4.38. Показания датчика давления 6.

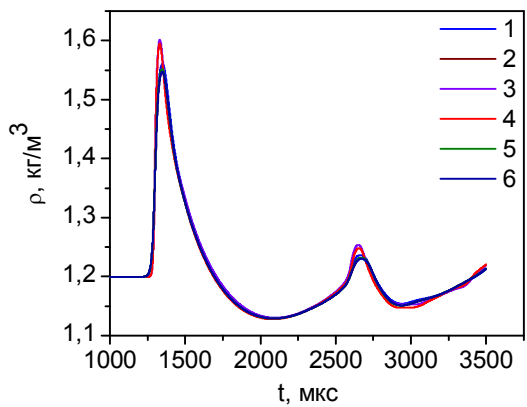

Рис. 4.39. Показания датчика плотности 1.

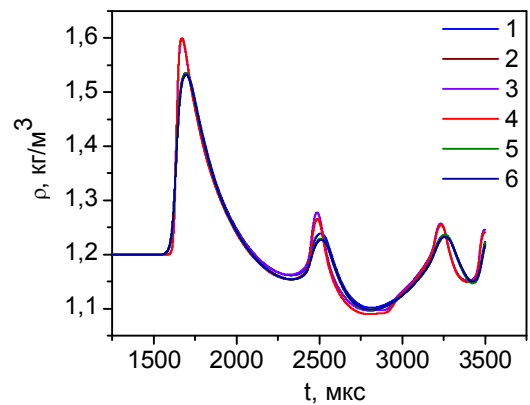

Рис. 4.40. Показания датчика плотности 2.

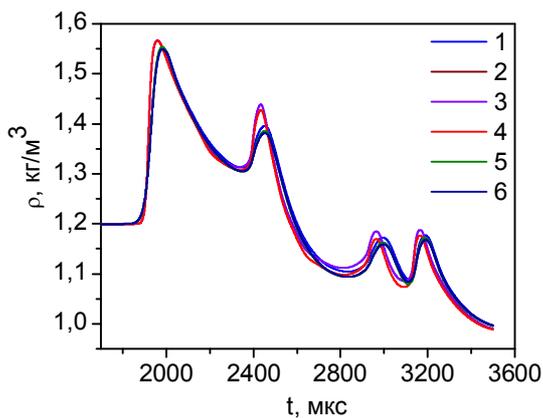

Рис. 4.41. Показания датчика плотности 3.

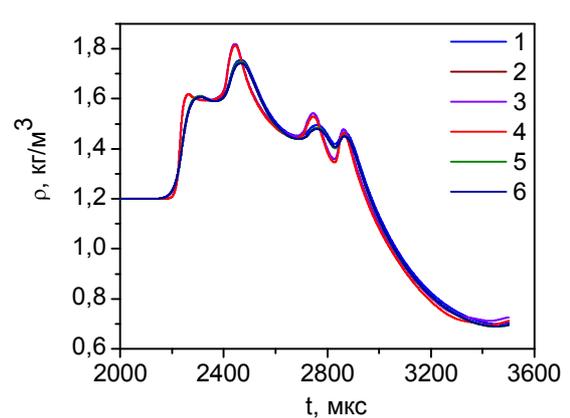

Рис. 4.42. Показания датчика плотности 4.

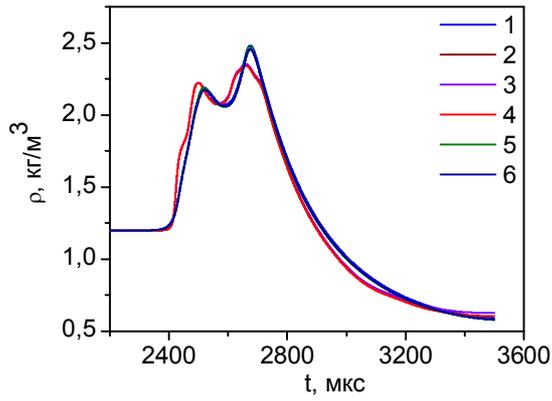

Рис. 4.43. Показания датчика плотности 5.

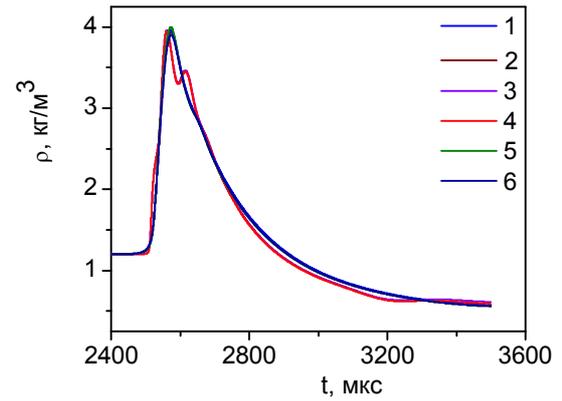

Рис.4.44. Показания датчика плотности 6.

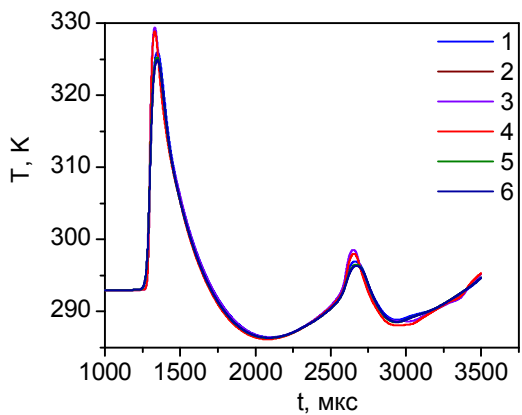

Рис. 4.45. Показания датчика температуры 1.

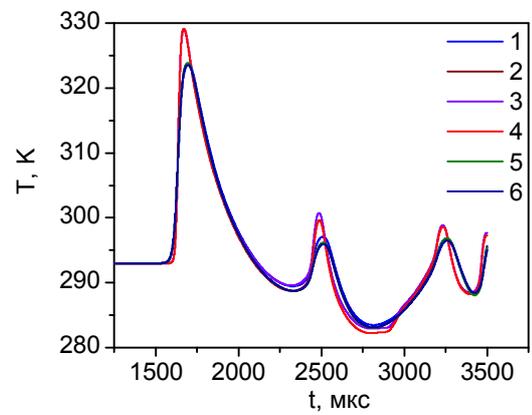

Рис.4.46. Показания датчика температуры 2.

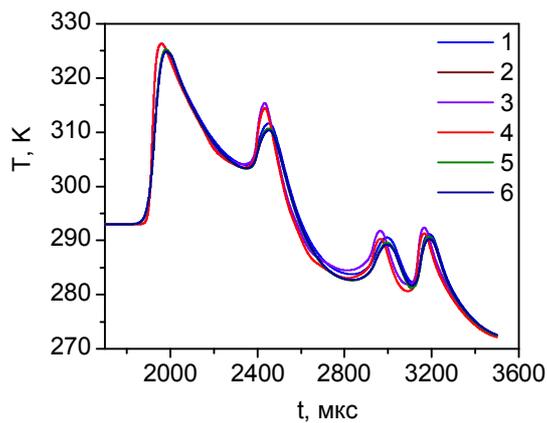

Рис. 4.47. Показания датчика температуры 3.

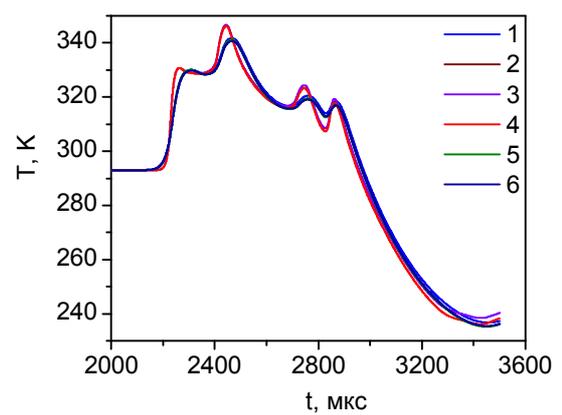

Рис.4.48. Показания датчика температуры 4.

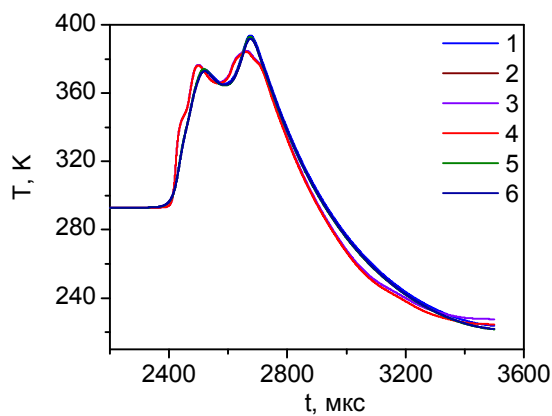 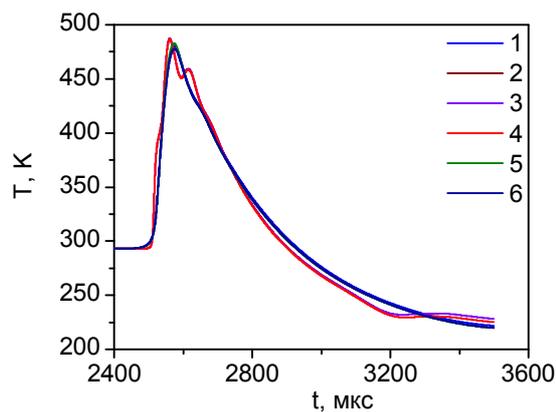

| Рис. 4.49. Показания датчика температуры 5. | Рис.4.50. Показания датчика температуры 6. |

### *4.3 Влияние мощности заряда и угла раскрытия конуса*

В данном разделе рассмотрены следующие факторы, влияющие на максимальное давление во время кумуляции ударной волны:

— изменение мощности инициирующего заряда,

— изменение угла раскрытия конуса.

На рис. 4.52 – 4.54 представлены профили давления на оси конуса в моменты времени 1300, 2600, 3500 мкс соответственно (вариант 4), демонстрирующие положительную и отрицательную фазы падающей и отраженной ударных волн. Кроме того был проведен расчет распространения ударной волны при энергии инициирования 19 кДж. Качественная картина течения совпадает с описанной выше для энергии 15 кДж (см. рис. 4.4 – 4.32).

Ниже представлены зависимости от времени максимального и минимального давления в области (рис. 4.55), а также давлений для каждого из шести датчиков (рис. 4.56–4.61). Для сравнения приведены результаты расчета случая 4 (см. таблицу 4.2), проведенного при тех же параметрах (геометрия, сетка, число Куранта), но при энергии инициирования 15 кДж. В таблице 4.3 приведены максимальные давления, зафиксированные для каждого датчика.

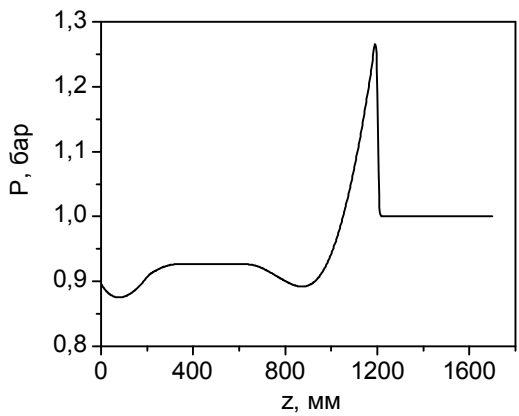
Рис. 4.52. Профиль давления на оси конуса (t = 1300мкс).

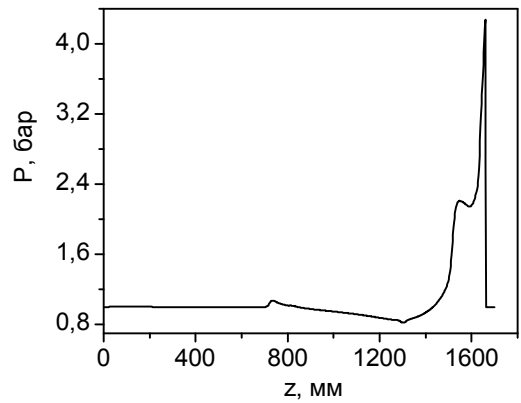
Рис. 4.53. Профиль давления на оси конуса (t = 2600мкс).

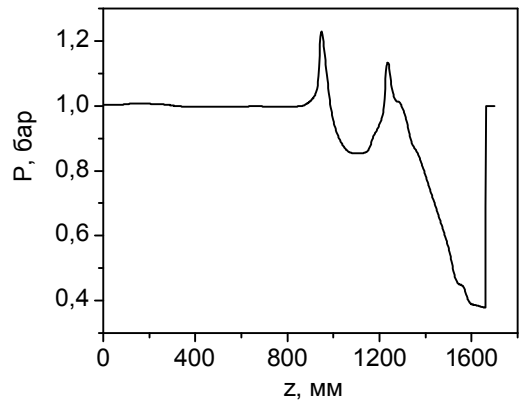
Рис. 4.54. Профиль давления на оси конуса (t = 3500мкс).

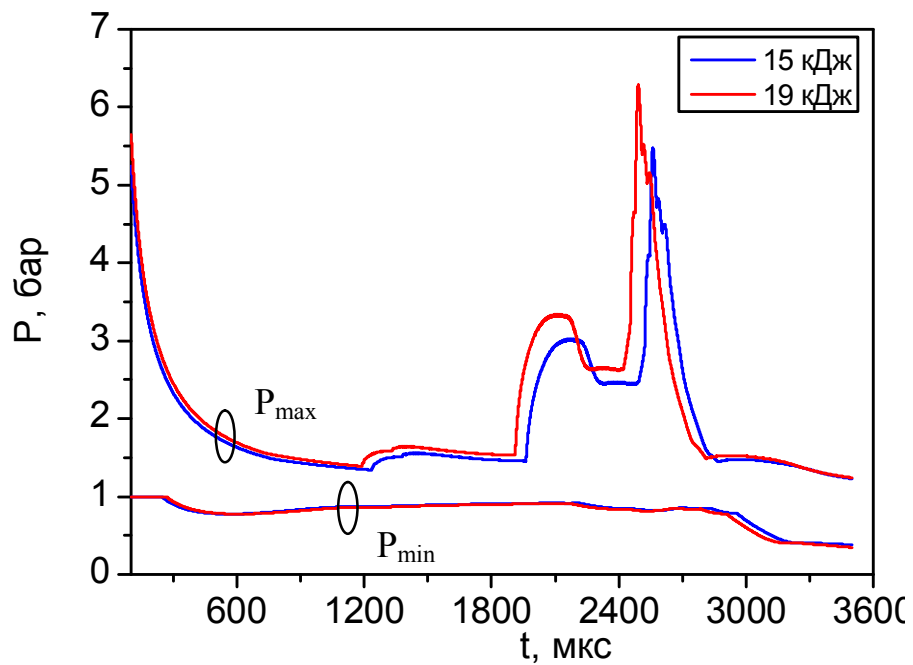
Рис.4.55. Зависимость максимального и минимального давления от времени для различных энергий инициирования

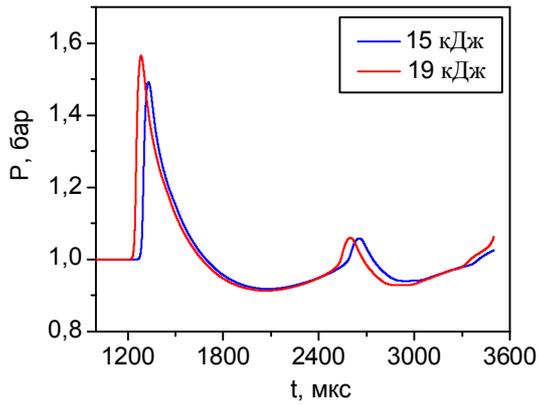

Рис. 4.56. Показания датчика давления 1.

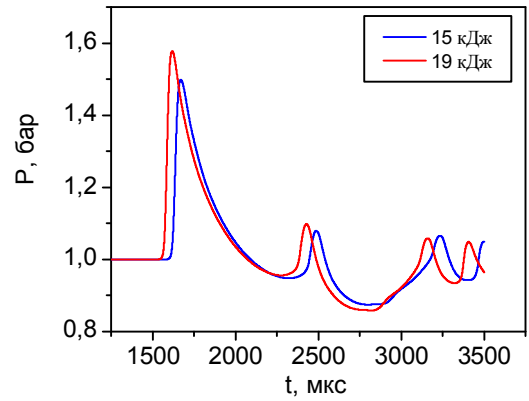

Рис. 4.57. Показания датчика давления 2.

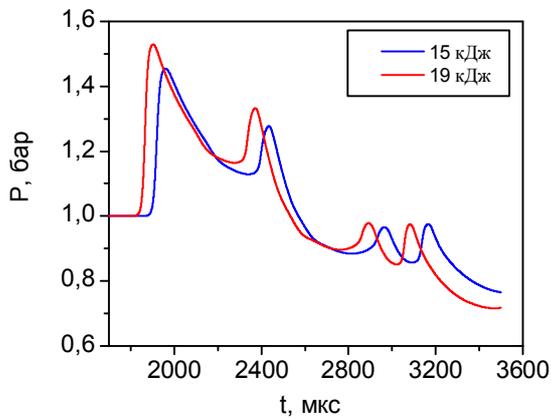

Рис. 4.58. Показания датчика давления 3.

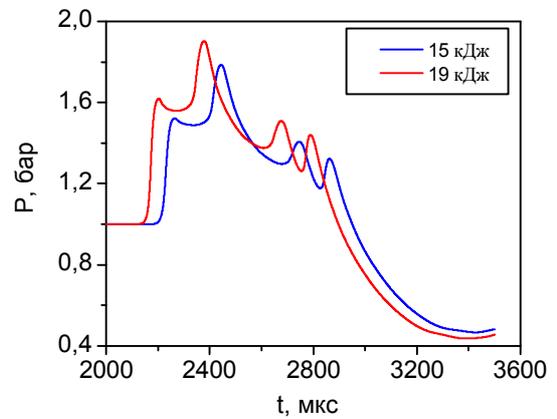

Рис. 4.59. Показания датчика давления 4.

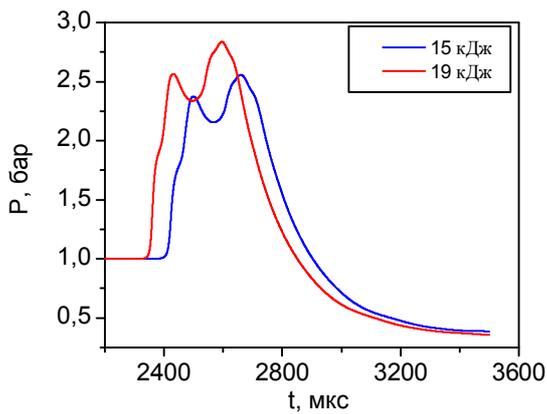

Рис. 4.60. Показания датчика давления 5.

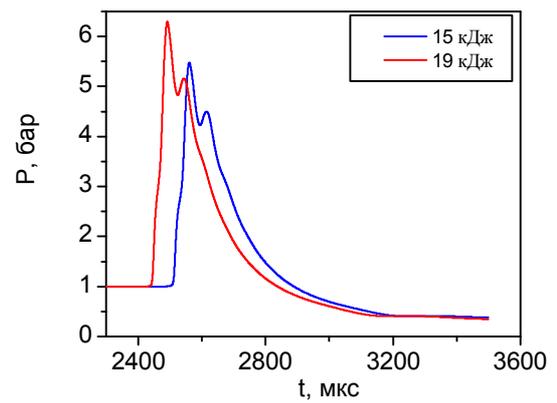

Рис. 4.61. Показания датчика давления 6.

Табл. 4.3. Максимальные давления.

| Энергия, кДж | $P_1$, бар | $P_2$, бар | $P_3$, бар | $P_4$, бар | $P_5$, бар | $P_6$, бар |
|---|---|---|---|---|---|---|
| 15 | 1.492 | 1.468 | 1.454 | 1.523 | 2.559 | 5.477 |
| 19 | 1.566 | 1.579 | 1.530 | 1.621 | 2.836 | 6.325 |

Для определения влияния угла раскрытия конуса на максимальное давление при кумуляции были проведены расчеты распространения ударной волны при углах, равных 20°, 25°, 35°, 40°, 45° (остальные параметры такие же, как и в расчете случая 4). Полученные результаты показывают, что наиболее сильная кумуляция происходит при угле раскрытия, равном 25° (рис. 4.62). Для уточнения полученного результата были проведены расчеты для углов раскрытия конуса, равных 22.5° и 27.5°. Результаты показывают, что наибольшее давление при кумуляции также приходится на 25° (рис. 4.63). Аналогичные расчеты были проведены при энергии инициирования 19 кДж. Качественная картина изменения максимального давления от угла раскрытия конуса совпадает с расчетами для энергии 15 кДж (рис. 4.64 – 4.66).

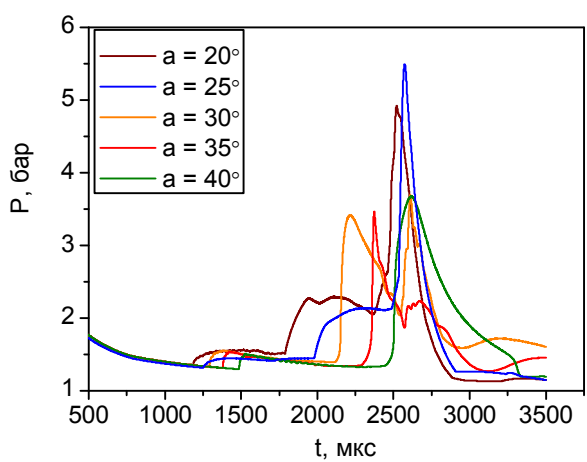 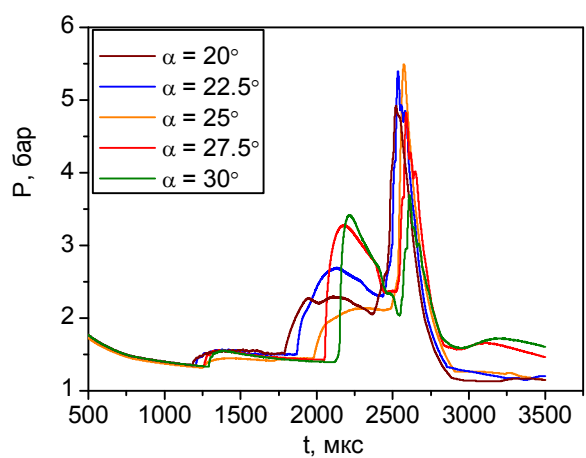

Рис. 4.62. Показания датчика максимального давления (15 кДж).

Рис. 4.63. Показания датчика максимального давления (15 кДж).

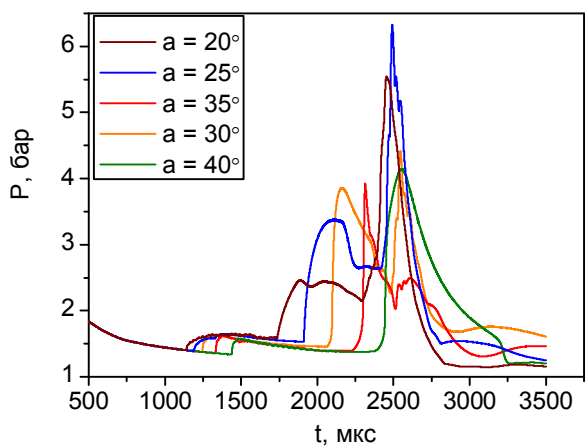 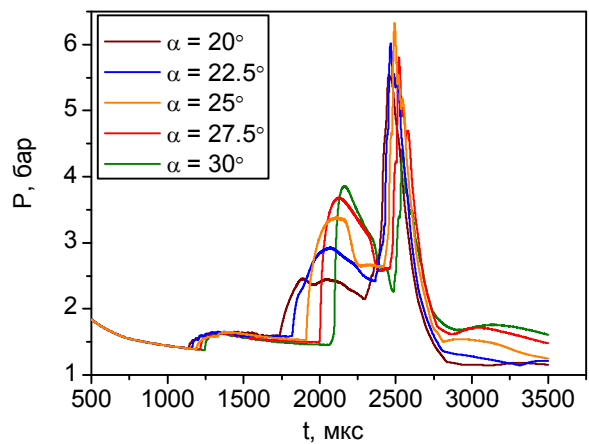

Рис. 4.64. Показания датчика максимального давления (19 кДж).

Рис. 4.65. Показания датчика максимального давления (19 кДж).

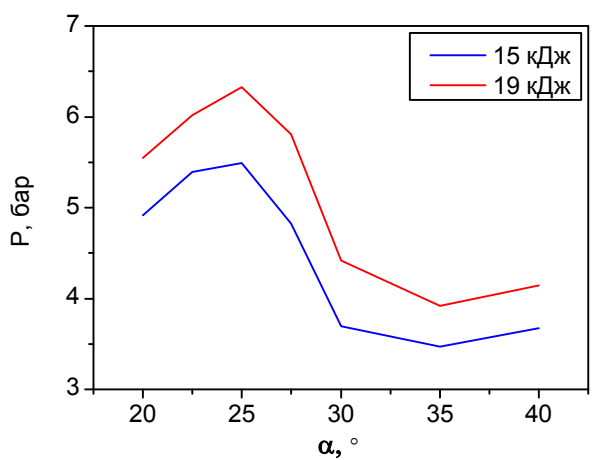

Рис. 4.66. Зависимость максимального давления при кумуляции УВ от угла раскрытия конуса.

## *4.4 Расчет детонации стехиометрической водородовоздушной смеси*

Были проведены численные расчеты распространения детонационной волны для варианта расчетной области, представленного на рис. 4.67. Красным цветом показана область течения, синим – заблокированная область. Размеры конуса и положения датчиков давления соответствовали размерам экспериментальной установки ОИВТ РАН (длина образующей конуса 600 мм, угол раствора составляет 25 градусов). Расчеты проводились для стехиометрической водородовоздушной смеси ($H_2 - 29.6\%, O_2 - 14.8\%$, $N_2 - 55.6\%$). Параметры расчетов представлены в табл. 4.4.

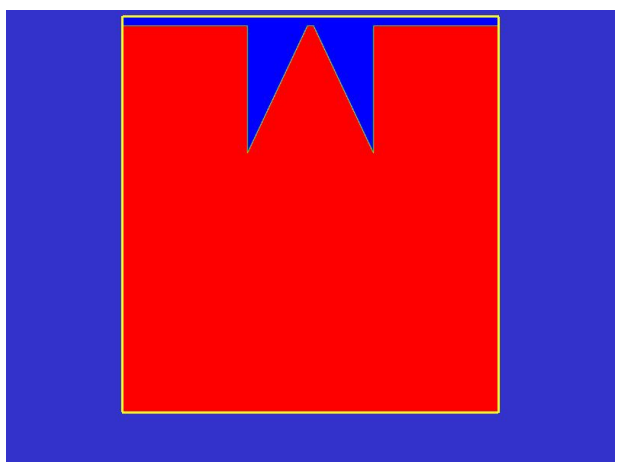

Рис. 4.67. Расчетная область.

Табл. 4.4. Параметры численных расчетов.

| № расчета | Число Куранта | Сетка | Угол раскрытия конуса | Инициирующий заряд, кДж |
|---|---|---|---|---|
| 1 | 0.25 | 400×800 | 25º | 15 |
| 2 | 0.25 | 400×800 | 25º | 19 |

На рис. 4.68 – 4.101 представлены стадии кумуляции детонационной волны в конусе (для расчета 2) и объемная скорость тепловыделения (W, Вт/м$^3$). Показаны поля давления в различные моменты времени после выделения энергии в источнике. Видно, что детонационная волна достигает угла конуса к моменту времени t = 250 мкс (см. рис. 4.77, 4.78). Взаимодействие детонационной волны с внутренней поверхностью конуса приводит к ее отражению и формированию волны, сходящейся к вершине конуса (рис. 4.78 – 4.86). В момент t = 400 мкс отраженные волны с противоположных сторон конуса «схлопываются» (рис. 4.88). Отражение от донышка конуса происходит в момент t = 450 мкс (рис. 4.92). Далее волна движется наружу, причем в верхней части конуса формируется область низкого давления (рис. 4.94 – 4.101).

На рис. 4.102 – 4.107 приведены показания датчиков давления 1 – 6 соответственно. На каждом графике построена зависимость давления от времени для расчета 2, представленного в таблице 4.4. На рис. 4.108 представлена зависимость максимального давления от времени для расчетов 1 и 2. Из этого рисунка видно, что после входа детонационной волны в конус (t = 275 мкс) различие в энергии инициирования практически не сказывается на максимальном давлении в полости. В таблице 4.5 приведены максимальные давления, зафиксированные для каждого датчика.

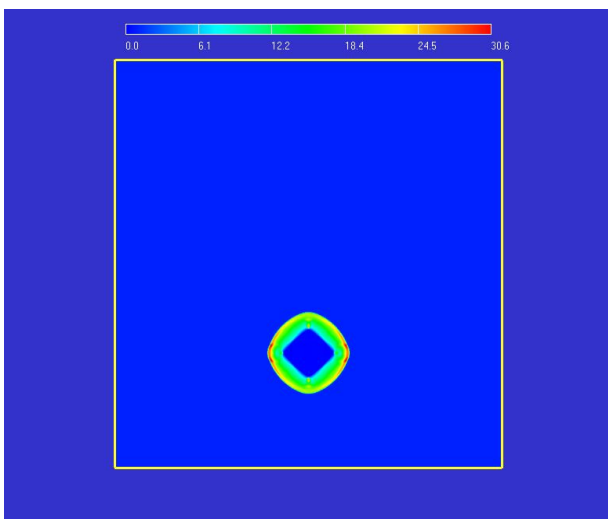 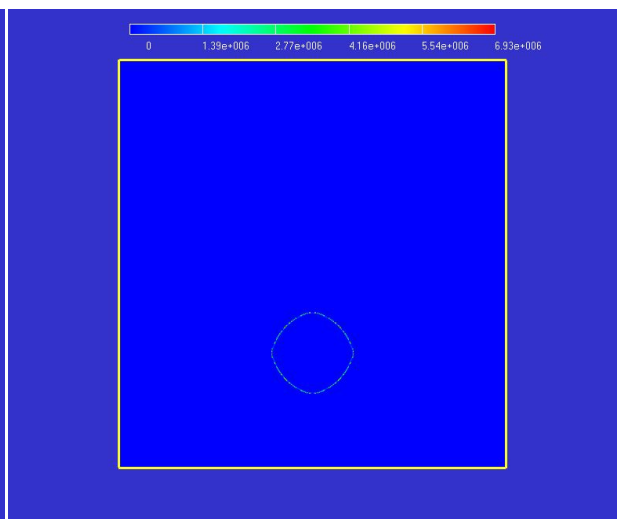

Рис. 4.68. Поле давления при кумуляции ДВ в момент t = 50 мкс.

Рис. 4.69. Объемная скорость тепловыделения W (t = 50 мкс).

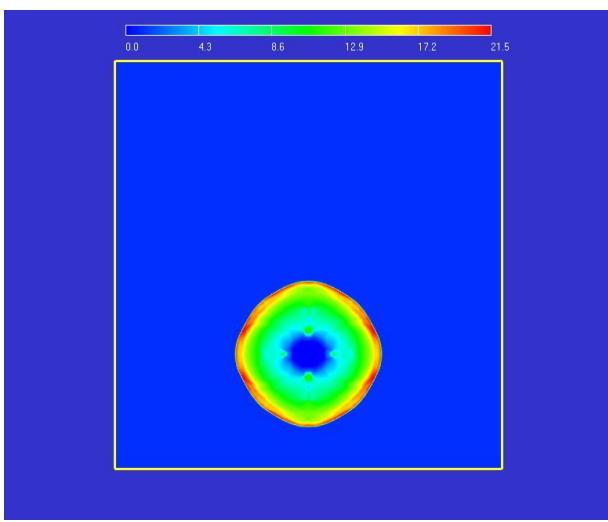 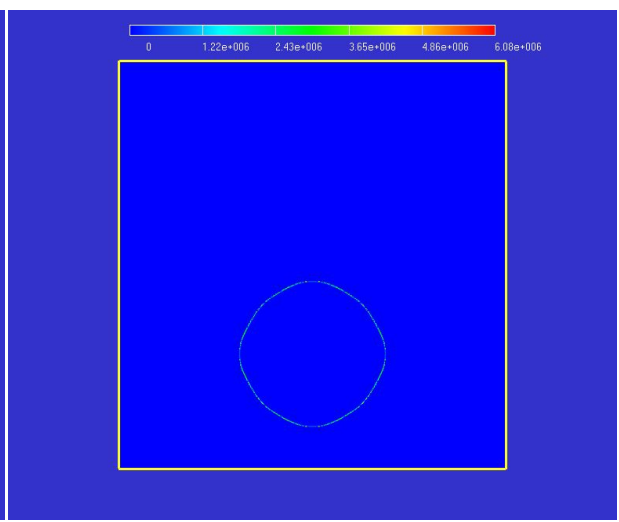

Рис. 4.70. Кумуляция ДВ (t = 100 мкс).

Рис. 4.71. Объемная скорость тепловыделения W (t = 100 мкс).

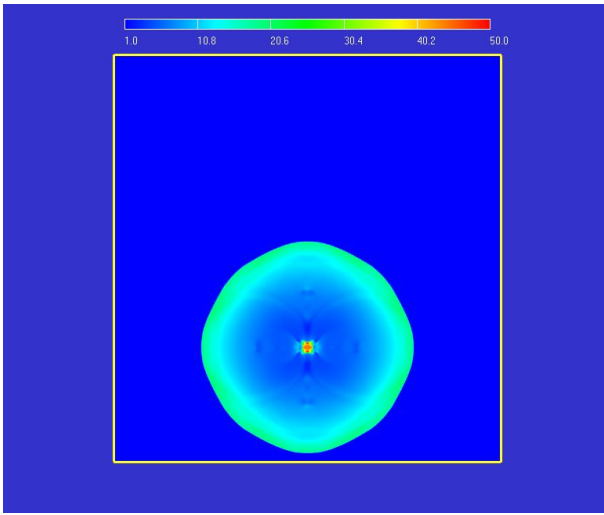 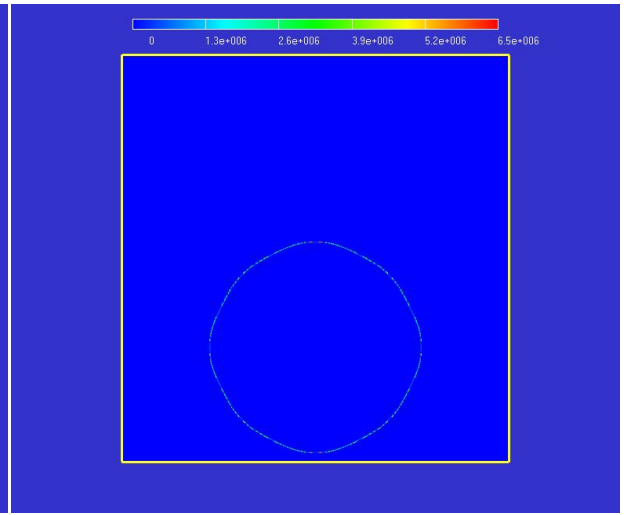

Рис. 4.72. Кумуляция ДВ (t = 150 мкс).

Рис. 4.73. Объемная скорость тепловыделения W (t = 150 мкс).

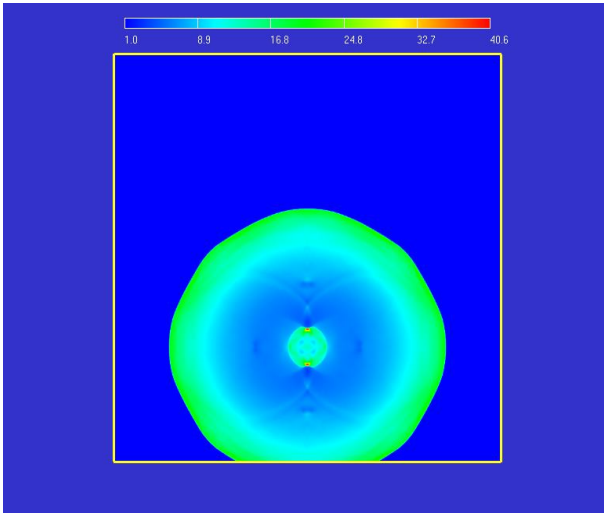 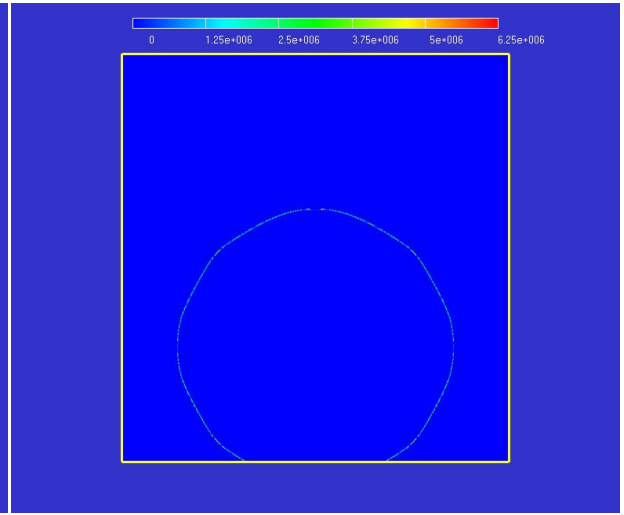

Рис. 4.74. Кумуляция ДВ (t = 200 мкс).

Рис. 4.75. Объемная скорость тепловыделения W (t = 200 мкс).

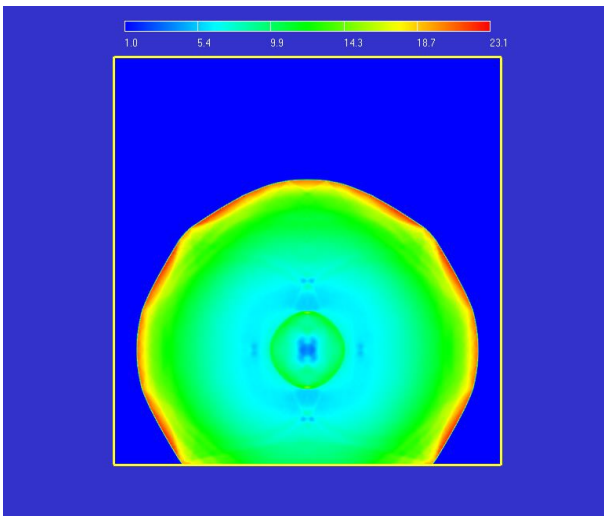 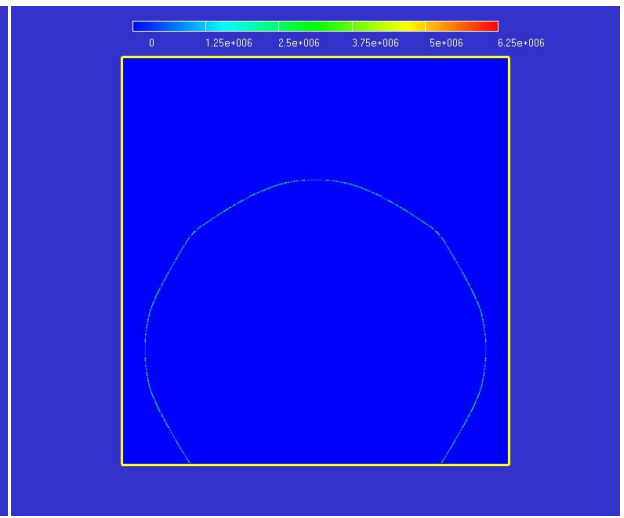

Рис. 4.76. Кумуляция ДВ (t = 250 мкс).

Рис. 4.77. Объемная скорость тепловыделения W (t = 250 мкс).

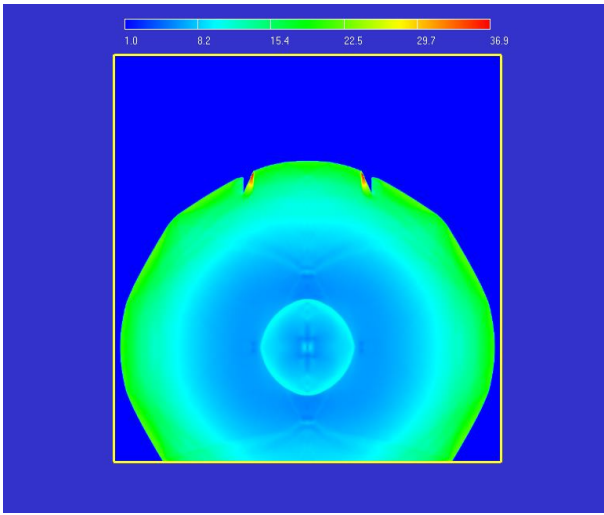

Рис. 4.78. Кумуляция ДВ (t = 275 мкс).

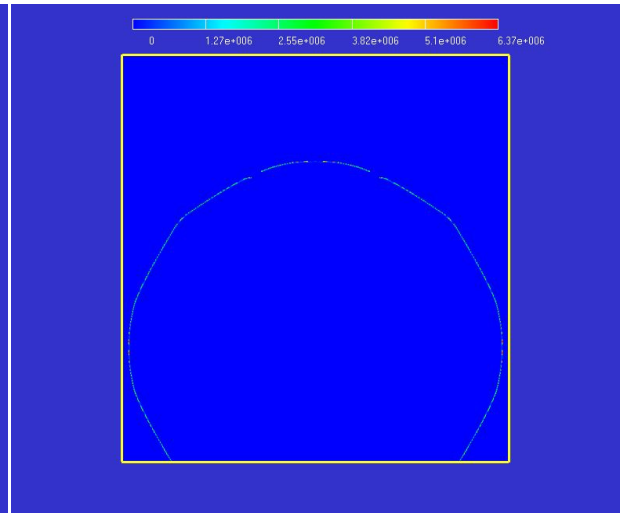

Рис. 4.79. Объемная скорость тепловыделения W (t = 275 мкс).

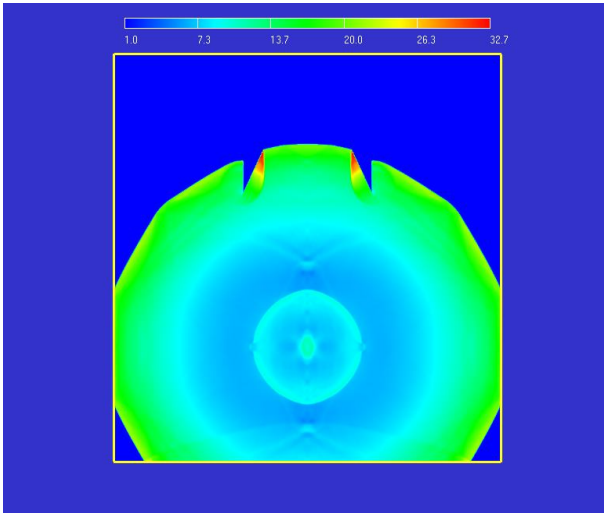

Рис. 4.80. Кумуляция ДВ (t = 300 мкс).

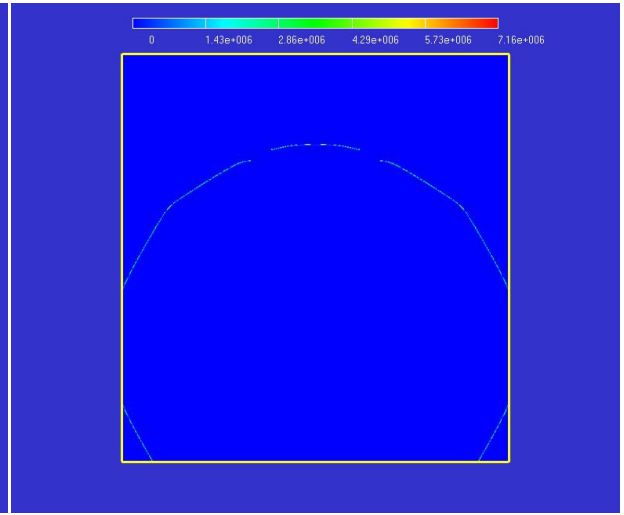

Рис. 4.81. Объемная скорость тепловыделения W (t = 300 мкс).

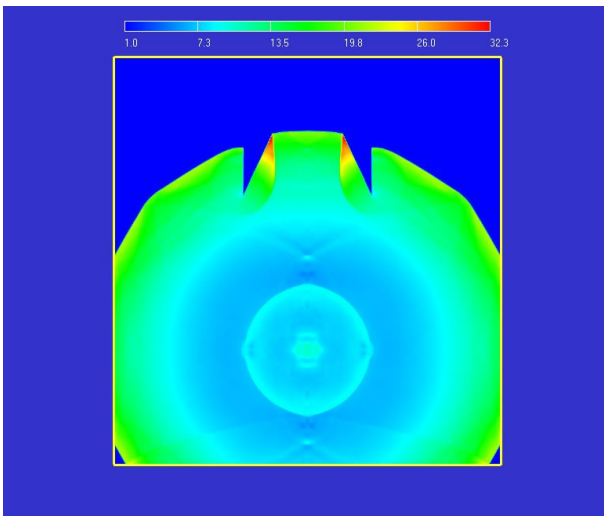

Рис. 4.82. Кумуляция УВ (t = 325 мкс).

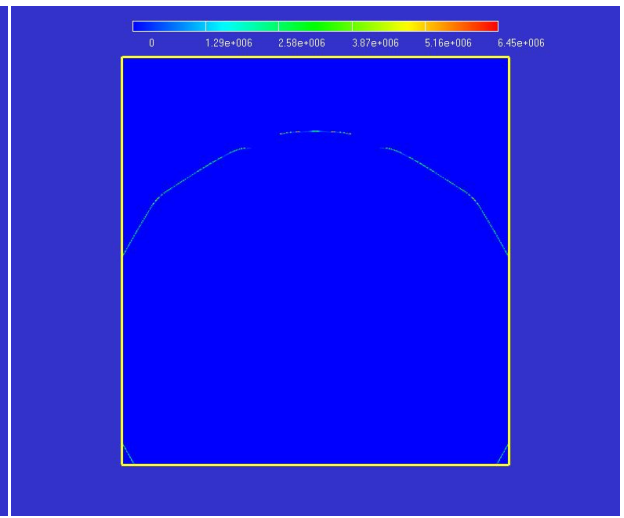

Рис. 4.83. Объемная скорость тепловыделения W (t = 325 мкс).

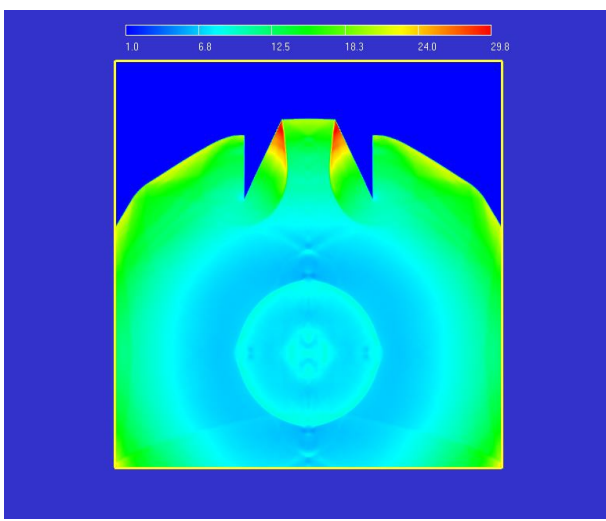 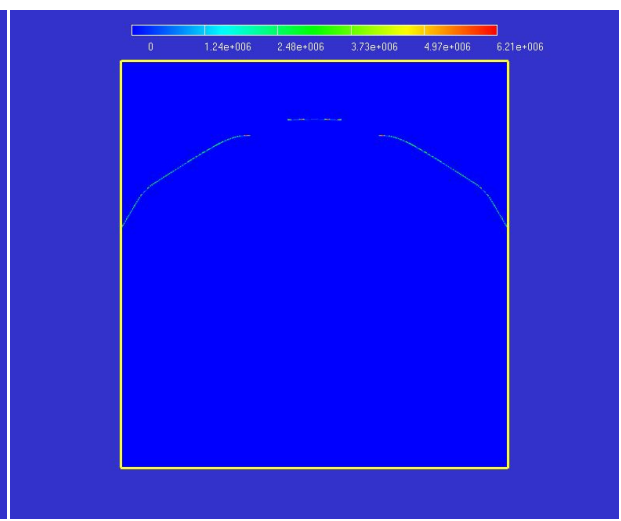

Рис. 4.84. Кумуляция ДВ (t = 350 мкс).

Рис. 4.85. Объемная скорость тепловыделения W (t = 350 мкс).

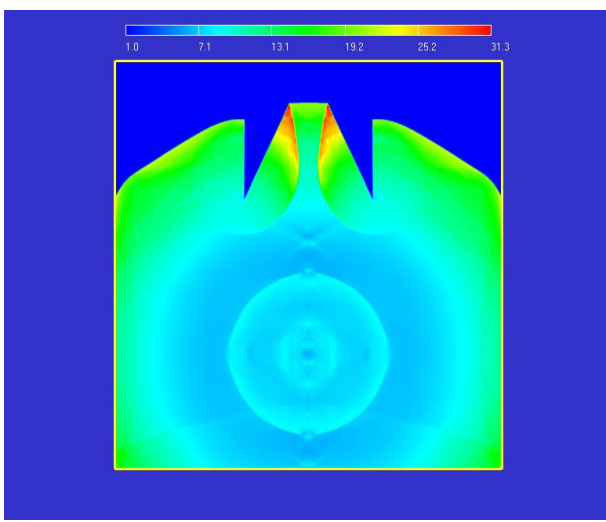 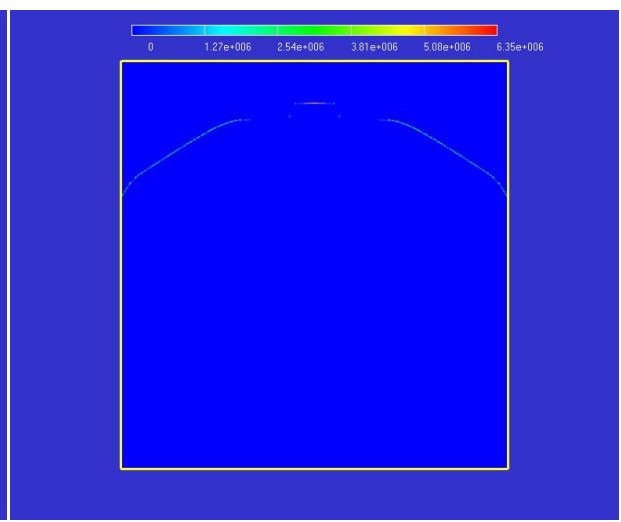

Рис. 4.86. Кумуляция ДВ (t = 375 мкс).

Рис. 4.87. Объемная скорость тепловыделения W (t = 375 мкс).

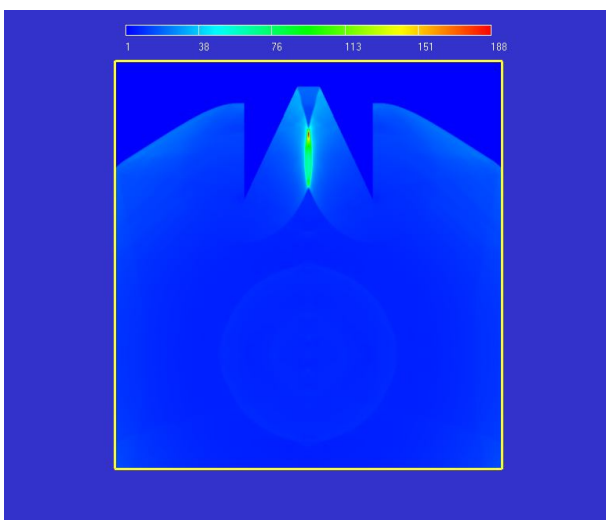 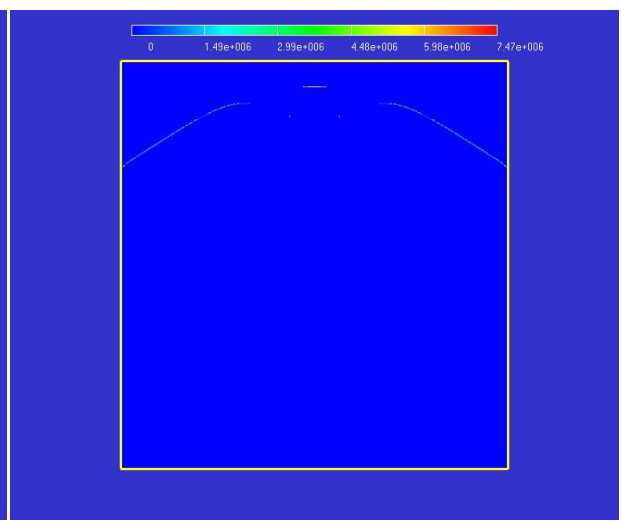

Рис. 4.88. Кумуляция ДВ (t = 400 мкс).

Рис. 4.89. Объемная скорость тепловыделения W (t = 400 мкс).

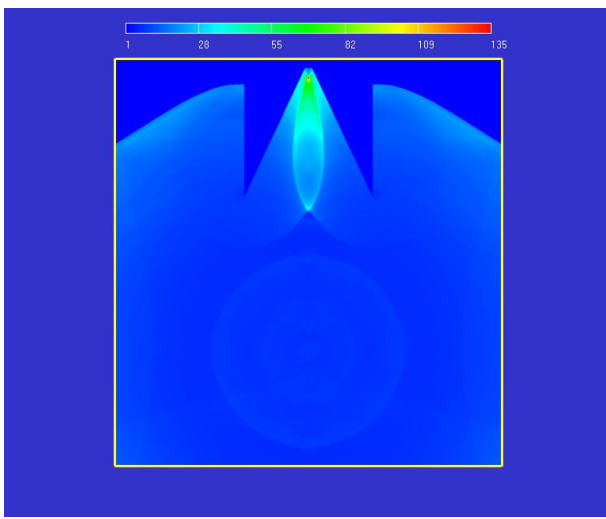 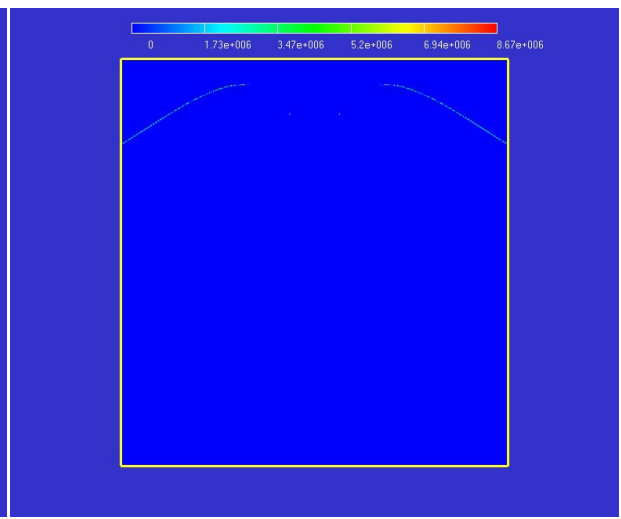

Рис. 4.90. Кумуляция ДВ (t = 425 мкс). | Рис. 4.91. Объемная скорость тепловыделения W (t = 425 мкс).

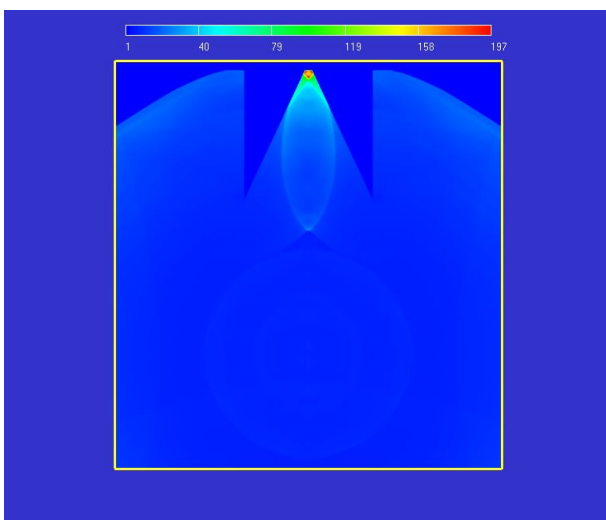 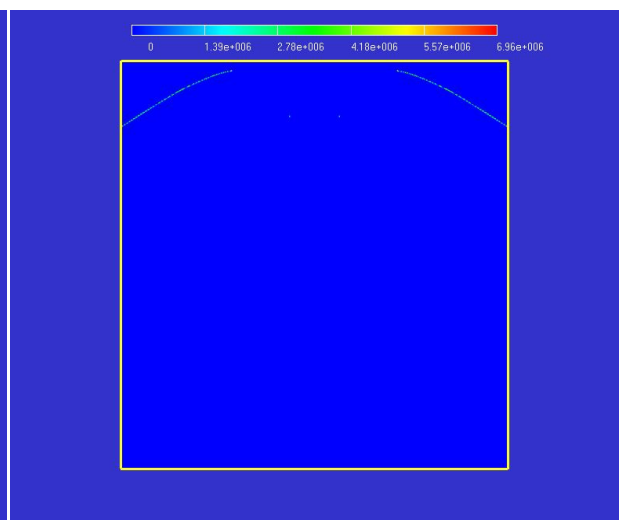

Рис. 4.92. Кумуляция ДВ (t = 450 мкс). | Рис. 4.93. Объемная скорость тепловыделения W (t = 450 мкс).

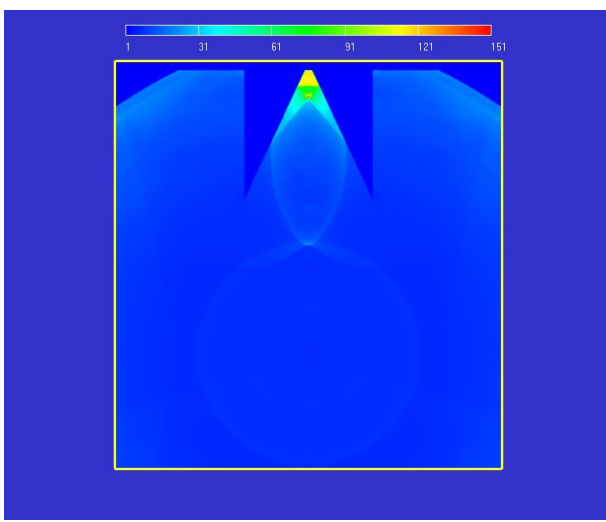 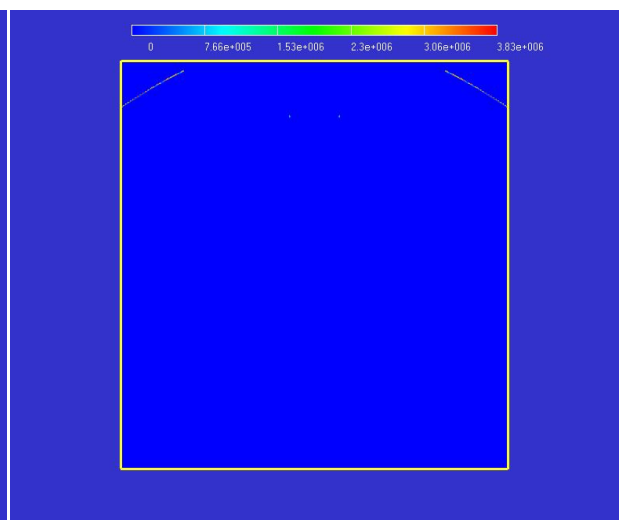

Рис. 4.94. Кумуляция ДВ (t = 475 мкс). | Рис. 4.95. Объемная скорость тепловыделения W (t = 475 мкс).

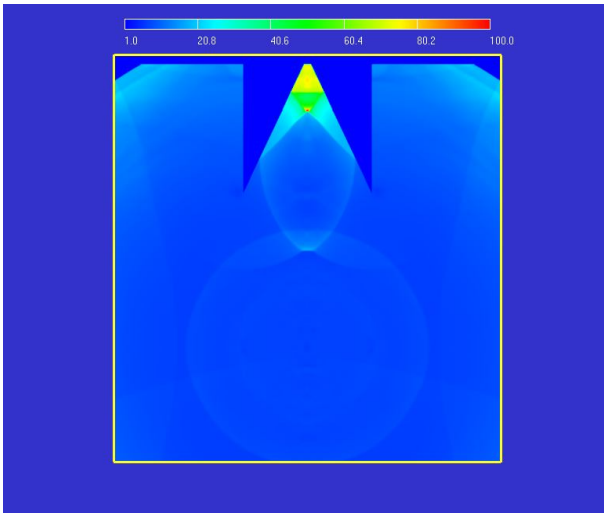 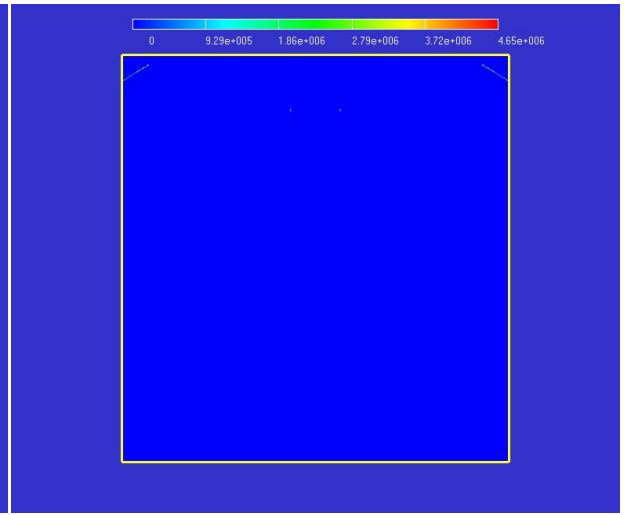

Рис. 4.96. Кумуляция ДВ (t = 500 мкс).

Рис. 4.97. Объемная скорость тепловыделения W (t = 500 мкс).

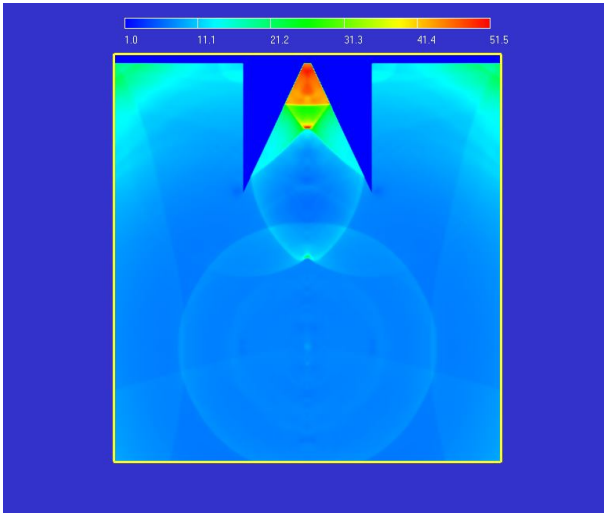 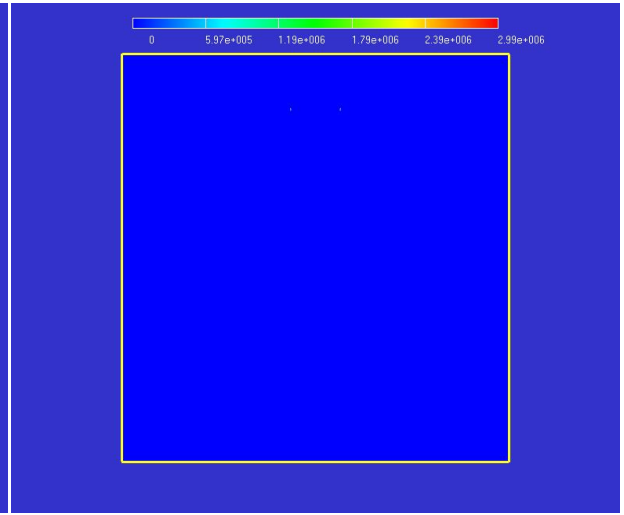

Рис. 4.98. Кумуляция ДВ (t = 525 мкс).

Рис. 4.99. Объемная скорость тепловыделения W (t = 525 мкс).

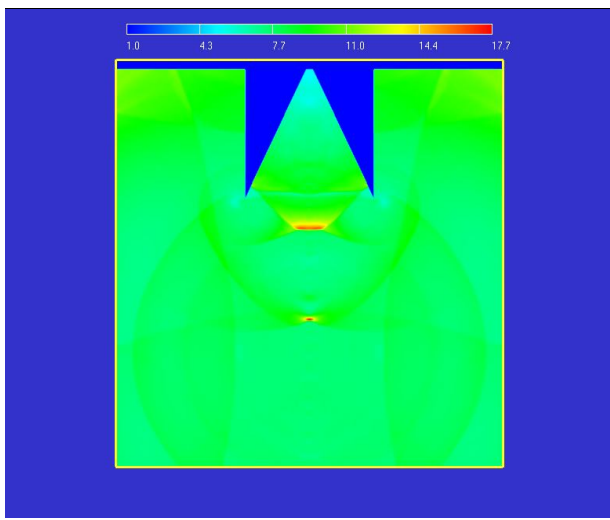 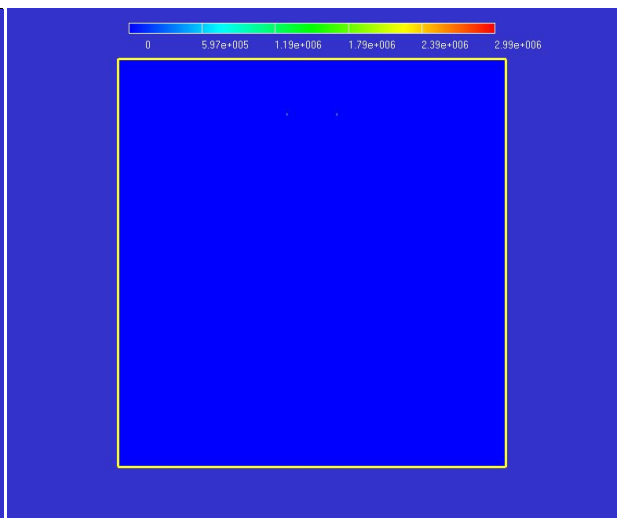

Рис. 4.100. Кумуляция ДВ (t = 675 мкс).

Рис. 4.101. Объемная скорость тепловыделения W (t = 675 мкс).

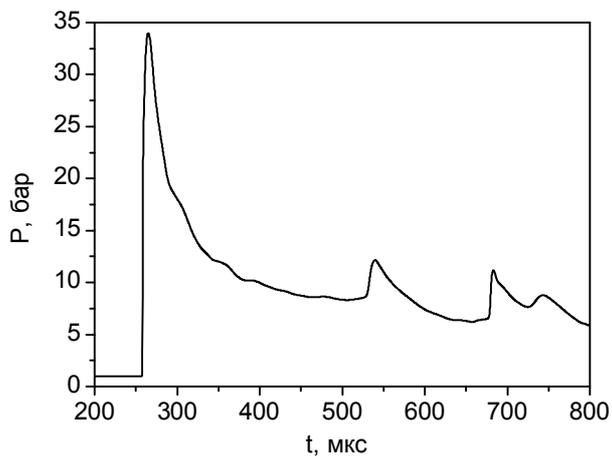

Рис. 4.102. Показания датчика давления 1.

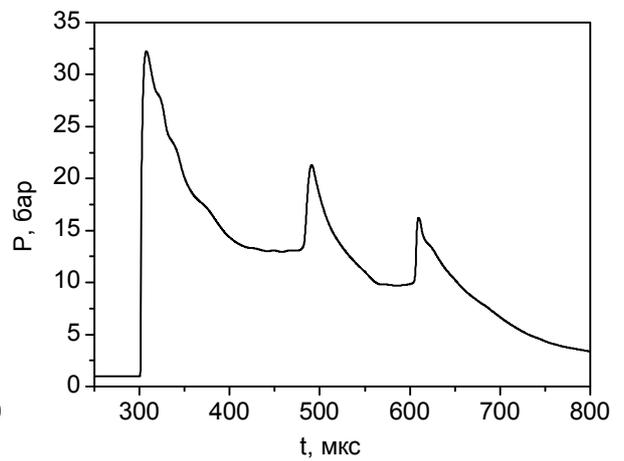

Рис. 4.103. Показания датчика давления 2.

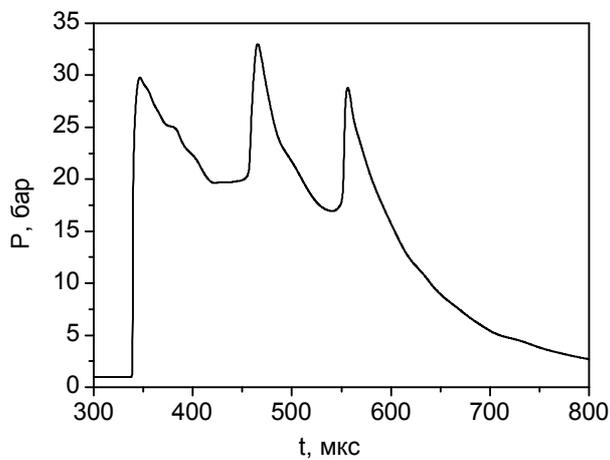

Рис. 4.104. Показания датчика давления 3.

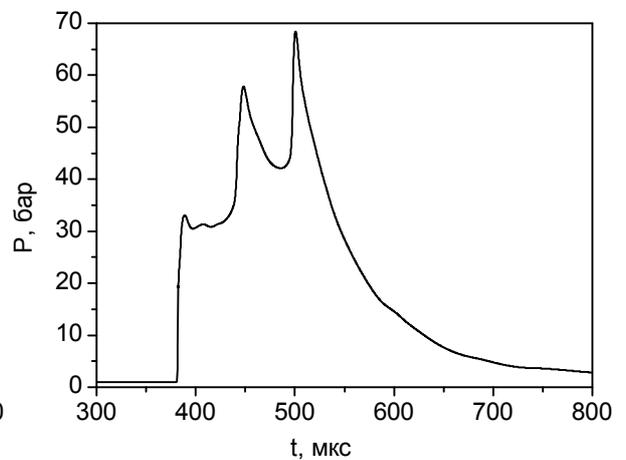

Рис. 4.105. Показания датчика давления 4.

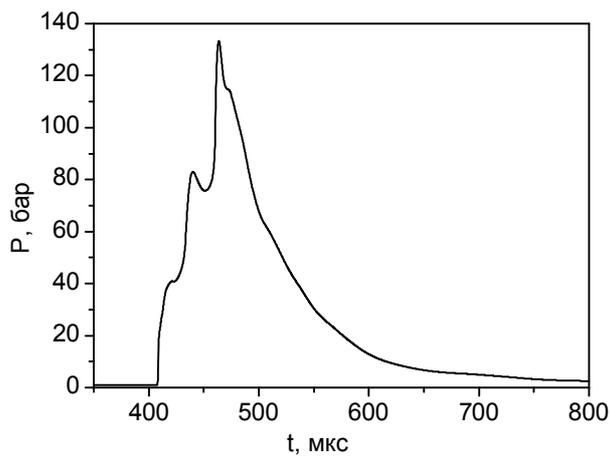

Рис. 4.106. Показания датчика давления 5.

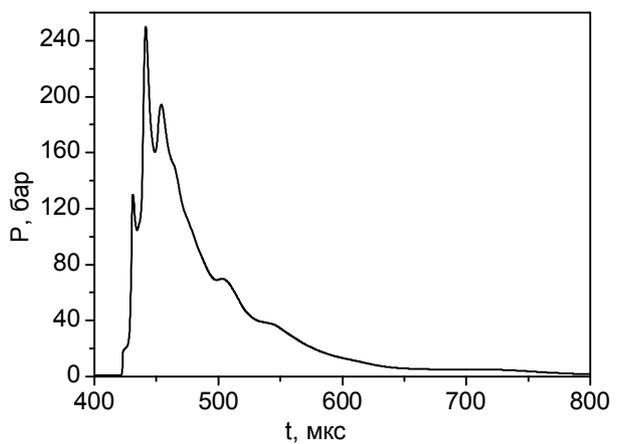

Рис. 4.107. Показания датчика давления 6.

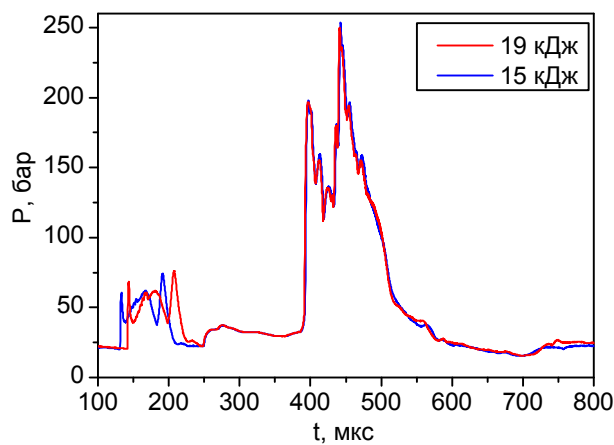

Рис. 4.108. Зависимость максимального давления от энергии.

Табл. 4.5. Максимальные давления для датчиков 1-6.

| Энергия, кДж | $P_1$, бар | $P_2$, бар | $P_3$, бар | $P_4$, бар | $P_5$, бар | $P_6$, бар |
|---|---|---|---|---|---|---|
| 15 | 34.11 | 32.07 | 33.18 | 69.6 | 135.31 | 253.41 |
| 19 | 33.99 | 32.23 | 33 | 68.45 | 133.43 | 250.04 |

Для определения влияния угла раскрытия конуса на максимальное давление при кумуляции были проведены расчеты распространения детонационной волны при тех же углах, что и в разделе 4.3 (остальные параметры такие же, как и для расчета 1 из табл. 4.4). Полученные результаты показывают, что наиболее сильная кумуляция происходит при угле раскрытия, равном 22.5° (рис. 4.108 – 4.109). Аналогичные результаты получены при энергии инициирования 19 кДж (рис. 4.110 – 4.111). Качественная картина изменения максимального давления от угла раскрытия конуса совпадает с расчетами для энергии 15 кДж в диапазоне углов раскрытия 20–27.5° (рис. 4.112).

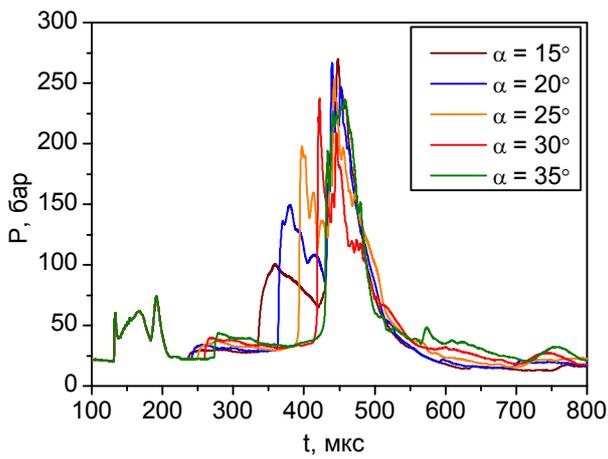

Рис. 4.108. Показания датчика максимального давления (15 кДж).

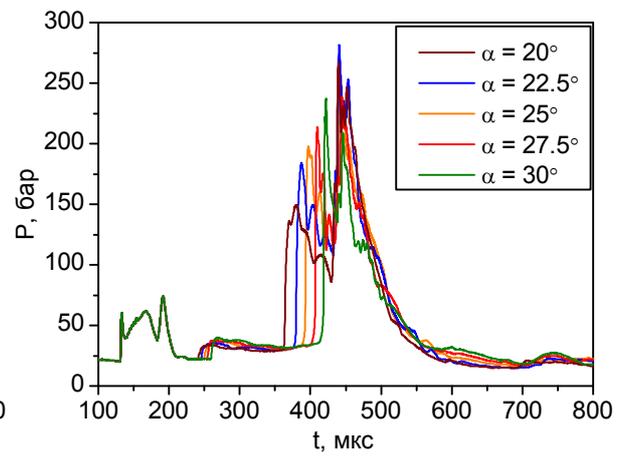

Рис.4.109. Показания датчика максимального давления (15 кДж).

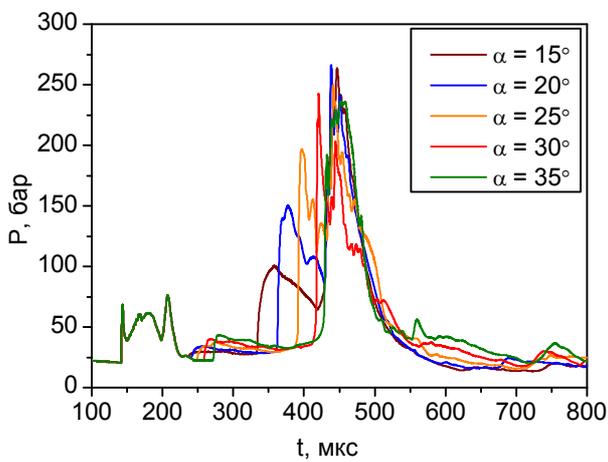

Рис. 4.110. Показания датчика максимального давления (19 кДж).

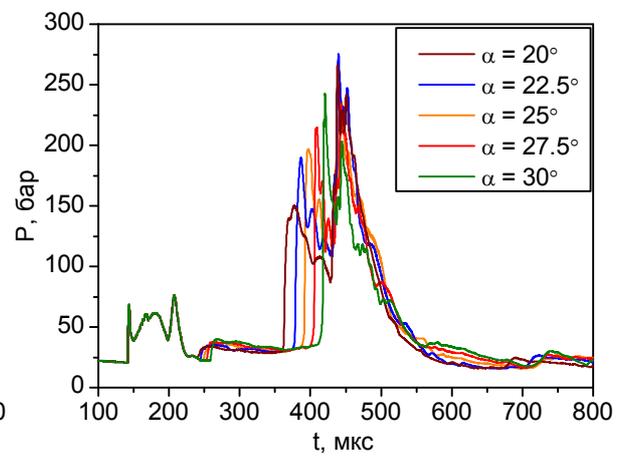

Рис.4.111. Показания датчика максимального давления (19 кДж).

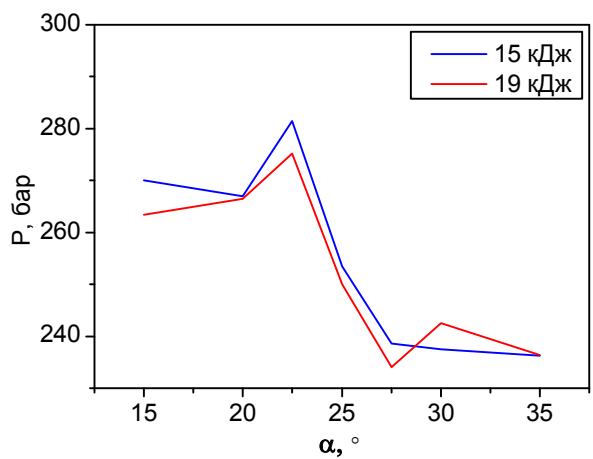

Рис. 4.112. Зависимость максимального давления при кумуляции ДВ от угла раскрытия конуса.

## 5 Выводы

Проведенные расчеты показывают, что используемая численная схема позволяет проводить расчеты ударных и детонационных волн в областях сложной геометрии. Качественная картина течения оказывается близкой для энергий инициирования 15 и 19 кДж. Наиболее сильная кумуляция ударной волны происходит при угле раскрытия конуса, равном 25°. В результате расчета распространения детонационной волны и сравнения с результатами экспериментов ОИВТ РАН получено хорошее качественное согласие:

1) временной последовательности развития процесса;
2) изменения давления вдоль образующей конуса: падение максимального давления от датчика 1 к датчику 3 и последующий рост на датчиках 4 – 6;
3) изменения давления в кумулирующемся потоке, а также значения давления и характер его изменения в возвратном течении после фокусировки в вершине.

Однако есть количественные отличия от экспериментов в результатах расчетов при энергии инициирования 19 кДж: максимумы давления для датчиков 1 – 6 наблюдаются в эксперименте раньше, чем при численном расчете, а величины этих давлений несколько меньше экспериментальных.

В дальнейшем планируется учет детальной химической кинетики, использование адаптивных расчетных сеток, учет показателей адиабаты для различных компонентов смеси.

## Литература